\documentclass{SciPost}

\hypersetup{
    colorlinks,
    linkcolor={red!50!black},
    citecolor={blue!50!black},
    urlcolor={blue!80!black}
}
\usepackage{algorithm,algpseudocode,mathtools}
\usepackage{orcidlink}
\usepackage{todonotes}
\usepackage{svg}
\usepackage{xcolor}
\usepackage{mathtools}
\usepackage[bitstream-charter]{mathdesign}

\usepackage{enumitem}
\usepackage{orcidlink}
\usepackage[acronym]{glossaries}
\newacronym{HS}{HS}{Hubbard-Stratonovich}
\newacronym{CDW}{CDW}{charge-density-wave}
\newacronym{BOW}{BOW}{bond-order-wave}
\newacronym{DQMC}{DQMC}{determinant quantum Monte Carlo}
\newacronym{QMC}{QMC}{Quantum Monte Carlo}
\newacronym{SSH}{SSH}{Su-Schrieffer-Heeger}
\newacronym{oSSH}{oSSH}{optical Su-Schrieffer-Heeger}
\newacronym{HMC}{HMC}{hybrid Monte Carlo}
\newacronym{HQMC}{HQMC}{hybrid quantum Monte Carlo}
\newacronym{2D}{2D}{two-dimensional}
\newacronym{1D}{1D}{one-dimensional}
\newacronym{eph}{$e$-ph}{electron-phonon}
\newacronym{ee}{$e$-$e$}{electron-electron}
\newacronym{FS}{FS}{Fermi surface}
\newacronym{MCMC}{MCMC}{Markov chain Monte Carlo}
\newacronym{LDP}{LDP}{local dispersionless phonon}
\newacronym{QHO}{QHO}{quantum harmonic oscillator}
\newacronym{EFA}{EFA}{exact Fourier acceleration}
\newacronym{EFA-HQMC}{EFA-HQMC}{exact Fourier acceleration hybrid quantum Monte Carlo}
\newacronym{BSS}{BSS}{Blankenbecler, Scalapino and Sugar}
\newacronym{DMRG}{DMRG}{density matrix renormalization group}
\newacronym{KPM}{KPM}{kernel polynomial method}
\newacronym{CG}{CG}{conjugate gradient}
\newacronym{diagMC}{diagMC}{diagrammatic Monte Carlo}
\newacronym{ED}{ED}{exact diagonalization}
\newacronym{FFT}{FFT}{Fast Fourier Transform}
\newacronym{MSCHK}{MSCHK}{minimal split checkerboard}
\newacronym{ST}{ST}{Suzuki-Trotter}
\newacronym{DEAC}{DEAC}{differential evolution for analytic continuation}
\newacronym{BZ}{BZ}{Brillouin zone}
\newacronym{RMS}{RMS}{root mean square}

\newacronym{smoqydqmc}{\href{https://github.com/SmoQySuite/SmoQyDQMC.jl.git}{\texttt{SmoQyDQMC.jl}}}{\href{https://github.com/SmoQySuite/SmoQyDQMC.jl.git}{\texttt{SmoQyDQMC.jl}}}
\glsunset{smoqydqmc}

\newacronym{smoqyelphqmc}{\href{https://github.com/SmoQySuite/SmoQyElPhQMC.jl.git}{\texttt{SmoQyElPhQMC.jl}}}{\href{https://github.com/SmoQySuite/SmoQyElPhQMC.jl.git}{\texttt{SmoQyElPhQMC.jl}}}
\glsunset{smoqyelphqmc}

\newacronym{smoqydeac}{\href{https://github.com/SmoQySuite/SmoQyDEAC.jl.git}{\texttt{SmoQyDEAC.jl}}}{\href{https://github.com/SmoQySuite/SmoQyDEAC.jl.git}{\texttt{SmoQyDEAC.jl}}}
\glsunset{smoqydeac}

\usepackage[acronym]{glossaries}
\usepackage{cleveref}

\usepackage{xcolor}
\usepackage{etoolbox}

\newtoggle{showchanges}
\togglefalse{showchanges}

\let\oldtextcolor\textcolor
\renewcommand{\textcolor}[2]{%
  \iftoggle{showchanges}{\oldtextcolor{#1}{#2}}{#2}%
}

\newcommand{\revise}[2][red]{%
  \iftoggle{showchanges}{\textcolor{#1}{#2}}{#2}%
}

\DeclareSymbolFont{usualmathcal}{OMS}{cmsy}{m}{n}
\DeclareSymbolFontAlphabet{\mathcal}{usualmathcal}

\fancypagestyle{SPstyle}{
\fancyhf{}
\lhead{\colorbox{scipostblue}{\bf \color{white} ~SciPost Physics Codebases }}
\rhead{{\bf \color{scipostdeepblue} ~Submission }}

\fancyfoot[C]{\textbf{\thepage}}
}

\DeclareMathOperator{\tr}{\text{tr}}

\begin{document}

\pagestyle{SPstyle}

\begin{center}{\Large \textbf{\color{scipostdeepblue}{
SmoQyElPhQMC.jl: An open-source Julia package for efficient and 
scalable quantum Monte Carlo simulations of electron-phonon coupled models
}}}\end{center}

\begin{center}\textbf{
Benjamin~Cohen-Stead\orcidlink{0000-0002-7915-6280}\textsuperscript{1,2$\star$},
James~Neuhaus\textsuperscript{1,2}
\orcidlink{0000-0001-6904-8510},
Kipton~Barros\textsuperscript{3}
\orcidlink{0000-0002-1333-5972}, 
Thomas~A.~Maier\orcidlink{0000-0002-1424-9996}\textsuperscript{4} and
Steven~Johnston\orcidlink{0000-0002-2343-0113}\textsuperscript{1,2$\dagger$}
}\end{center}

\begin{center}
{\bf 1} Department of Physics and Astronomy, The University of Tennessee, Knoxville, Tennessee 37996, USA
\\
{\bf 2} Institute of Advanced Materials and Manufacturing, The University of Tennessee, Knoxville, Tennessee 37996, USA\\
{\bf 3} Theoretical Division and CNLS, Los Alamos National Laboratory, Los Alamos, New Mexico 87545, USA 
\\
{\bf 4} Computational Sciences and Engineering Division, Oak Ridge National Laboratory, Oak Ridge, Tennessee, 37831-6494, USA
\\[\baselineskip]
$\star$ \href{mailto:bcohenst@utk.edu}{\small bcohenst@utk.edu}\,,\quad
$\dagger$ \href{mailto:sjohn145@utk.edu}{\small sjohn145@utk.edu}
\end{center}

\section*{\color{scipostdeepblue}{Abstract}}
\textbf{\boldmath{
We introduce version 1.0 of the \gls*{smoqyelphqmc} package, an open-source Julia code for performing scalable quantum Monte Carlo simulations of electron-phonon coupled model Hamiltonians. \gls*{smoqyelphqmc} is built upon the \gls*{smoqydqmc} codebase and implements improved versions of the algorithms presented in [B. Cohen-Stead \textit{et al}., Phys. Rev. E 105, 065302 (2022)] to enable \revise{(near-linear) scalable} simulations of a broad class of uncorrelated $e$-ph models both in system size and inverse temperature. By extending the functionality of the flexible scripting interface introduced in \gls*{smoqydqmc}, the \gls*{smoqyelphqmc} package continues to allow users to adapt it to different workflows and interface with other software packages in the Julia ecosystem. The code for this package can be downloaded from our GitHub repository at \url{https://github.com/SmoQySuite/SmoQyElPhQMC.jl} or installed using the Julia package manager.  The online documentation, including examples, can be obtained from our documentation page at  \url{https://smoqysuite.github.io/SmoQyElPhQMC.jl/stable/}.}
}

\vspace{\baselineskip}

\noindent\textcolor{white!90!black}{%
\fbox{\parbox{0.975\linewidth}{%
\textcolor{white!40!black}{\begin{tabular}{lr}%
  \begin{minipage}{0.6\textwidth}%
    {\small Copyright attribution to authors. \newline
    This work is a submission to SciPost Physics Codebases. \newline
    License information to appear upon publication. \newline
    Publication information to appear upon publication.}
  \end{minipage} & \begin{minipage}{0.4\textwidth}
    {\small Received Date \newline Accepted Date \newline Published Date}%
  \end{minipage}
\end{tabular}}
}}
}


\vspace{10pt}
\noindent\rule{\textwidth}{1pt}
\tableofcontents
\noindent\rule{\textwidth}{1pt}
\vspace{10pt}


\section{Introduction}\label{sec:intro}
\subsection{Overview and Scope}\label{sec:overview}
This paper introduces the \gls*{smoqyelphqmc} package, a user-friendly open-source Julia package for performing efficient and scalable \gls*{QMC} simulations of uncorrelated \gls*{eph} coupled models. The package is part of the SmoQy Suite and extends the functionality of the \gls*{smoqydqmc} package~\cite{Cohen-Stead2024SmoQyDQMCjl} by implementing a modified version of the algorithm introduced in Ref.~\cite{Cohen-Stead2022Fast}. In doing so, the package enables \revise{scalable} \gls*{QMC} simulations (both in system size $N$ and inverse temperature $\beta = 1/k_\mathrm{B}T$, where $k_\mathrm{B}$ is the Boltzmann constant) of a large set of generalized spin-symmetric \gls*{eph} models \revise{in the absence of} any Hubbard interactions. 

The \gls*{smoqyelphqmc} package follows a similar design philosophy as other codes in the SmoQy suite, and expands the scripting interface used by the \gls*{smoqydqmc} package. This document focuses on outlining the features of the \gls*{smoqyelphqmc} package and discusses the algorithms that are unique to it, including any modifications and improvements on the original algorithms described in Ref.~\cite{Cohen-Stead2022Fast}. It also serves as a citable document when using this code for research. For a detailed discussion of the components related to the \gls*{smoqydqmc} package, we refer the reader to Ref.~\cite{Cohen-Stead2024SmoQyDQMCjl} and its associated documentation. This document is not intended to serve as a detailed user manual. Instead, we maintain extensive \href{https://smoqysuite.github.io/SmoQyElPhQMC.jl/stable/}{online documentation} for \gls*{smoqyelphqmc}, with docstrings, examples, and other reference material to be added over the package's lifetime.

\subsection{Background and Motivation}\label{sec:background}
The \gls*{eph} interaction is ubiquitous in condensed matter systems, and is responsible for a host of scientifically and technologically important phenomena. It plays a central role in determining the transport properties of many materials~\cite{mahan2000many, Giri2020electron, Kazemian2026electron}, it renormalizes the electron~\cite{Engelsberg1963coupled, Lanzara2001evidence, Cuk2005review, Grothe2013quantifying, Mazzola2013kinks, You2025diverse} and phonon dispersion \revise{relations}~\cite{Taylor1963theory, Renker1973observation, Koenig1864kohn, Piscanec2004kohn, Nocerino2023, Korshunov2024phonon}, and drives the formation of (bi)polaron quasi-particles in the strong coupling limit~\cite{devreese2015frohlich, Franchini2021polarons, dai2025polarons}. This interaction can also lead to various broken symmetry states including conventional superconductivity~\cite{Bardeen1957theory, Carbotte1990properties, marsiglio2001electron}, \gls*{CDW}~\cite{Grunner1988dyanmics, Xuetao2015classification, Alidoosti2021charge}, orbital ~\cite{Souliou2016soft, Johnston2014charge}, valence bond~\cite{Weber2021valence, Gotz2022valence, MalkarugeCosta2024Kekule}, and even magnetic~\cite{Cai2021antiferromagnetism, Feng2022phase, Cai2025quantum} orders. More recently, \gls*{eph} interactions have even been linked to the properties of some topological states of matter~\cite{Wan2014turning, Moller2017typeII, Heid2017electron}. 

The study of \gls*{eph} interactions has a long and distinguished history (see Ref.~\cite{dai2025polarons} for a recent review), and numerical simulations have played a prominent role in developing our understanding of this interaction. Migdal-Eliashberg theory, for example, provides a powerful general framework for calculating \gls*{eph} self-energies and for predicting superconducting and competing \gls*{CDW} instabilities~\cite{Scalapino1969, marsiglio2001electron, Marsiglio2020eliashberg}. However, it generally breaks down in either the antiadiabatic ($\hbar\Omega/E_\mathrm{F}\ge 1$) or strong-coupling ($\lambda\ge 1$) limits~\cite{Bauer2011quantitative, Esterlis2018Breakdown, Yuzbashyan2022breakdown}, where polaron formation and other lattice instabilities become important.  
(Here $E_\mathrm{F}$ is the Fermi energy, $\hbar\Omega$ denotes the typical phonon energy, and $\lambda$ is a dimensionless measure of the \gls*{eph} coupling strength.) 
\revise{First-principles} methods for modeling \gls*{eph} interactions and static polaron properties have also become quite advanced in recent years~\cite{Savrasov1996electron, Giustino2017electronphonon, Franchini2021polarons, dai2025polarons}. Nevertheless, these 
approaches are generally unable to address polaron dynamics and other phenomena that require treating the lattice beyond the Born-Oppenheimer approximation. Instead, progress in tackling the strong-coupling regime while treating the electronic and lattice subsystems on equal footing has largely relied on applying nonperturbative many-body methods to model Hamiltonians. 

Diagrammatic Monte Carlo has long been the method of choice for computing the properties of lattice polarons in the dilute limit~\cite{Prokofev1998polaron, Mishchenko2000diagrammatic, Greitemann2018lecture}, and has recently been extended to higher carrier concentrations for selected models~\cite{Mishchenko2014diagrammatic}. Variational approaches like the momentum average approximation~\cite{Goodvin2006Greens, Covaci2009Polaron} are another class of powerful methods for treating \gls*{eph} models in the dilute limit. \Gls*{QMC}~\cite{Scalettar1989Competition, Macridin2006synergistic, Hardikar2007phase, Johnston2013Determinant, Weber2015excitation, Esterlis2018Breakdown, Costa2018Phonon, Karakuzu2018solution, Li2019electronic, Costa2020phase, Bradley2021Superconductivity, Nosarzewski2021superconductivity, Xing2021quantum, Weber2021valence, Cai2021antiferromagnetism, Feng2022phase, MalkarugeCosta2023Comparative, Bradley2023Charge, MalkarugeCosta2024Kekule, TanjaroonLy2023Comparative} and \gls*{DMRG}~\cite{Jeckelmann1998density, Bursill1998phase, Jeckelmann1999metal, Fehske2008metallicity, Ejima2010dmrg, Nocera2014interplay, Jansen2020finite, Banerjee2023Groundstate, Tang2023traces, Nocera2023electron, Zhao2023onedimensional, Thomas2025theory, Kovac2025signiture, Banerjee2025Spectral} have also been applied broadly to study various \gls*{eph} models with great success. However, due to technical limitations, many of these methods have been limited \revise{to models} with relatively high-energy phonons, where $\hbar\Omega$ is comparable to the electron hopping. Wavefunction-based methods like \gls*{ED} and \gls*{DMRG}, for example, have to overcome large Fock spaces needed to describe systems with small phonon energies $\hbar\Omega$. While these limitations may be partially mitigated via a clever choice of basis~\cite{Cataudella2004variational, Wang2020zero}, they have generally limited these calculations to small clusters, \gls*{1D} systems, systems in the dilute limit, and coupling to high-energy modes. (The momentum-average method has also recently been extended to the adiabatic regime but remains restricted to dilute systems~\cite{Carbone2021numerically}.) Similarly, auxiliary field \gls*{QMC} methods relying on performing local updates for sampling often suffer from long autocorrelation times~\cite{Hohenadler2008} and ergodicity~\cite{Scalettar1991ergodicity, Johnston2013Determinant} issues when sampling low-energy phonons. As a result, numerical simulations of \gls*{eph} models have been broadly limited in the accessible system size and/or addressing the challenging adiabatic regime $\hbar\Omega/E_\mathrm{F} \ll 1$ at finite carrier concentrations relevant to many materials. 

A variety of improved sampling methods have been developed to address these challenges in \gls*{QMC} simulations, including self-learning Monte Carlo~\cite{Liu2017self, Li2019Accelerating, Chen2018Symmetryenforced}, Langevin~\cite{Batrouni2019Langevin, Cohen-Stead2020Langevin}, and \gls*{HMC}~\cite{Duane1987Hybrid, Neal2011MCMC}. In particular, \gls*{HMC} algorithms are particularly powerful for efficiently sampling continuous fields in both classical Monte Carlo and \gls*{QMC} simulations, and have recently emerged as an indispensable tool for simulating \gls*{eph} models~\cite{Cohen-Stead2022Fast, Beyl2018Revisiting}. For example, \gls*{HMC} has already been applied to  \gls*{DQMC} simulations of \gls*{eph} models, where \revise{it grants} access to low-energy optical\cite{Cohen-Stead2023Hybrid, Bradley2023Charge, MalkarugeCosta2024Kekule} and acoustic\cite{MalkarugeCosta2023Comparative} phonon branches. Notably, \gls*{HMC} methods can be combined with several other algorithmic advances~\cite{Cohen-Stead2022Fast, Ostmeyer2025Minimal} to achieve \revise{near-linear $O(\beta \mathcal{N})$ scaling as the adiabatic limit is approached} in systems absent an \gls*{ee} interaction. 

The \gls*{smoqyelphqmc} package provides capabilities for performing large-scale \gls*{HMC} simulations for generic \revise{real} tight-binding Hamiltonians with generalized \gls*{eph} interactions. The code leverages updated versions of the algorithms outlined in Ref.~\cite{Cohen-Stead2022Fast} to efficiently sample the phonon fields for the same class of generalized \gls*{eph} models supported by the \gls*{smoqydqmc} package. These include 
\begin{enumerate}[itemsep=-1mm]
    \item{arbitrary lattices and bases \revise{in zero, one, two, and three dimensions};}
    \item{fully momentum-dependent \gls*{eph} interactions, including long-range Holstein- and \gls*{SSH}-like couplings;}
    \item{coupling to multiple phonon branches, either via the same or different microscopic coupling mechanisms;}
    \item{low-energy optical and acoustic phonon branches; }
    \item{nonlinear \gls*{eph} interactions and anharmonic lattice potentials up to fourth order in the atomic displacements;  }
    \item{dynamical tuning of the chemical potential to achieve a targeted density $\langle n\rangle$~\cite{Miles2022Dynamical};  }
    \item{spatial disorder in any Hamiltonian parameter; and}
    \item{specialized support for efficiently measuring a wide range of common observables on arbitrary lattice geometries.}
\end{enumerate}

\subsection{Relevant links, documentation, and reporting}\label{sec:documentation}

\gls*{smoqyelphqmc}'s source code and its associated auxiliary packages can be found on the \href{https://github.com/SmoQySuite/SmoQyElPhQMC.jl}{SmoQy Suite's GitHub page}~\cite{SmoQySuite}. The package is registered in \revise{Julia's} \href{https://github.com/JuliaRegistries/General.git}{General registry}, and can also be installed using the Julia package manager by issuing the command 
\begin{verbatim}
julia> ]
pkg> add SmoQyElPhQMC
\end{verbatim}
Note that the \gls*{smoqydqmc} package, along with its low-level packages, will be installed as a dependency. 

\revise{\gls*{smoqyelphqmc}'s documentation can be found \href{https://smoqysuite.github.io/SmoQyDQMC.jl/stable/}{online}~\cite{SmoQyElPhQMC_docs}.} As with other codes in the SmoQy Suite, this documentation includes a full API and a growing number of example scripts. Any issues or questions related to the code can be submitted through its GitHub issues page.

\section{Supported Hamiltonians}

\gls*{smoqyelphqmc} supports a subset of \revise{spin-symmetric} Hamiltonians available in the \gls*{smoqydqmc} package but excludes Hubbard-like interactions due to limitations in its underlying sampling algorithms. This section discusses how the various Hamiltonian terms are parameterized, which is very similar to the parameterization used by \gls*{smoqydqmc}. Throughout, we use bold roman indices (e.g., $\mathbf{i}, \mathbf{j}, \dots$) to index the unit cells in the lattice, and Greek symbols (e.g. $\nu, \gamma, \dots$) to index the orbitals and phonon modes within each unit cell. We also normalize units such that $\hbar = 1$ and denote the total number of orbitals as $\mathcal{N} = N \cdot n$, where $N$ is the total number of unit cells in the lattice and $n$ is the number of orbitals in each unit cell. Similarly, the total number of phonon modes in the lattice is $\mathcal{N}_\text{ph} = N \cdot n_\text{ph}$, where $n_\text{ph}$ is the number of phonon modes per unit cell.

It is convenient to partition the full Hamiltonian as 
\begin{equation}
    \hat{\mathcal{H}} = \hat{\mathcal{U}} + \hat{\mathcal{V}} + \hat{\mathcal{K}},
\end{equation}
where $\hat{\mathcal{U}}$ describes the non-interacting lattice (phonon) degrees of freedom, and $\hat{\mathcal{V}}$ and $\hat{\mathcal{K}}$ describe the total electron potential and kinetic energies, respectively. In general, both $\hat{\mathcal{V}}$ and $\hat{\mathcal{K}}$ will depend on the dynamical lattice coordinates and thus include \gls*{eph} interactions that are diagonal and off-diagonal in the orbital basis, respectively. The former terms arise in Holstein- and Fr{\"o}hlich-like models\revise{,} while the latter arise in \gls*{SSH}- or Peierls-like models. 

The non-interacting lattice Hamiltonian is represented as the sum of three terms
\begin{equation}\label{eq:U_ph}
    \hat{\mathcal{U}} = \hat{\mathcal{U}}_\text{qho} + \hat{\mathcal{U}}_\text{anh} + \hat{\mathcal{U}}_\text{disp}.
\end{equation}
The first term
\begin{equation}\label{eq:U_qho}
    \hat{\mathcal{U}}_\text{qho} = \sum_\mathbf{i,\nu} \left[ \frac{1}{2M_{\mathbf{i},\nu}} \hat{P}_{\mathbf{i},\nu}^2 + \frac{1}{2} {M}_{\mathbf{i},\nu} \Omega_{\mathbf{i},\nu}^2 \hat{X}^2_{\mathbf{i},\nu} \right]
\end{equation}
defines bare \gls*{QHO} modes $\nu$ within each unit cell $\mathbf{i}$ with position \revise{$\hat{X}_{\mathbf{i},\nu}$} and momentum $\hat{P}_{\mathbf{i},\nu}$ operators for each mode $\nu$. Here, $\Omega_{\mathbf{i},\nu}$ and $M_{\mathbf{i},\nu}$ are the frequency and mass of the phonon mode and $K_{\mathbf{i},\nu} = {M}_{\mathbf{i},\nu} \Omega_{\mathbf{i},\nu}^2$
is the corresponding force constant. The second term 
\begin{equation}\label{eq:U_anh}
\hat{\mathcal{U}}_\text{anh} = \sum_{\mathbf{i},\nu} \frac{1}{24} M_{\mathbf{i},\nu} \Omega_{\revise{\mathrm{a}},\mathbf{i},\nu}^2 \hat{X}_{\mathbf{i},\nu}^4
\end{equation}
allows one to introduce a quartic anharmonic term in the phonon potential energy whose strength is controlled by $\Omega_{\revise{\mathrm{a}},\mathbf{i},\nu}$. Lastly, the third term
\begin{equation}\label{eq:U_disp}
    \hat{\mathcal{U}}_\text{disp} = \revise{\frac{1}{2}}\sum_{\substack{\mathbf{i},\nu \\ \mathbf{j},\eta}} \frac{M_{\mathbf{i},\nu}M_{\mathbf{j},\eta}}{M_{\mathbf{i},\nu}+M_{\mathbf{j},\eta}} \left[ \tilde{\Omega}_{(\mathbf{i},\nu),(\mathbf{j},\eta)}^2(\hat{X}_{\mathbf{i},\nu}-\xi_{(\mathbf{i},\revise{\nu}),(\mathbf{j},\eta)}\hat{X}_{\mathbf{j},\eta})^2 + \frac{1}{12}\tilde{\Omega}_{\revise{\mathrm{a}},(\mathbf{i},\nu),(\mathbf{j},\eta)}^2(\hat{X}_{\mathbf{i},\nu}-\xi_{(\mathbf{i},\revise{\nu}),(\mathbf{j},\eta)}\hat{X}_{\mathbf{j},\eta})^4 \right]
\end{equation}
introduces couplings between pairs of phonon modes in the lattice and $\xi_{(\mathbf{i},\revise{\nu}),(\mathbf{j},\eta)} = \pm 1$ is a parameter controlling the character of the dispersive coupling. Here, the sum runs over all unit cells $\mathbf{i} \ (\mathbf{j})$ and phonon modes $\nu \ (\eta)$ in each unit cell\revise{, and the factor of a half out front is to avoid double counting}. The parameterization given in Eqs.~\eqref{eq:U_qho}-\eqref{eq:U_disp} is flexible enough to allow one to simulate systems with multiple phonon branches, including combinations of dispersionless (Einstein) and dispersive acoustic and optical phonon modes. 

The total electron kinetic energy is conveniently subdivided as
\begin{equation}
    \hat{\mathcal{K}} = \hat{\mathcal{K}}_0 + \hat{\mathcal{K}}_\text{ssh}.
\end{equation}
The first term provides a tight-binding description of the non-interacting electron kinetic energy
\begin{equation}
    \hat{\mathcal{K}}_0 = -\sum_\sigma \sum_{\substack{(\mathbf{i},\gamma) \\ (\mathbf{j},\rho)}} t_{(\mathbf{i},\gamma),(\mathbf{j},\rho)} \left[
         \hat{c}_{\sigma,\mathbf{i},\gamma}^{\dagger} \hat{c}_{\sigma,\mathbf{j},\rho}^{\phantom\dagger}
        + \hat{c}_{\sigma,\mathbf{j},\rho}^{\dagger}
        \hat{c}_{\sigma,\mathbf{i},\gamma}^{\phantom\dagger}
    \right],
\end{equation}
where $\hat{c}_{\sigma,\mathbf{i},\gamma}^{\dagger} \ (\hat{c}_{\sigma,\mathbf{i},\gamma}^{\phantom\dagger})$ is the spin-$\sigma$ electron creation (annihilation) operator for orbital $\gamma$ in the unit cell $\mathbf{i}$ and $t_{(\mathbf{i},\gamma),(\mathbf{j},\rho)}$ is the hopping amplitude to go from orbital $\rho$ in unit cell $\mathbf{j}$ to orbital $\gamma$ in unit cell $\mathbf{i}$. The second term
\begin{equation}\label{eq:K_ssh}
    \hat{\mathcal{K}}_\text{ssh} = \sum_\sigma \sum_{\substack{(\mathbf{i},\gamma, \nu) \\ (\mathbf{j},\rho, \eta)}}\sum_{n=1}^{4} \alpha_{n,(\mathbf{i},\gamma,\nu),(\mathbf{j},\rho,\eta)} \left(\hat{X}_{\mathbf{i},\nu} - \hat{X}_{\mathbf{j},\eta}\right)^n
    \left[
         \hat{c}_{\sigma,\mathbf{i},\gamma}^{\dagger} \hat{c}_{\sigma,\mathbf{j},\rho}^{\phantom\dagger}
        + \hat{c}_{\sigma,\mathbf{j},\rho}^{\dagger}
        \hat{c}_{\sigma,\mathbf{i},\gamma}^{\phantom\dagger}
    \right]
\end{equation}
introduces an \gls*{SSH}-like \gls*{eph} coupling whereby the phonon positions modulate the hopping amplitudes between orbitals in the lattice. The parameterization given in Eq.~\eqref{eq:K_ssh} allows one to include interactions up to fourth order in the displacements, with the strength of the coupling set by $\alpha_{n,(\mathbf{i},\gamma,\nu),(\mathbf{j},\rho,\eta)}$. 

Similarly, the total electron potential energy is conveniently expressed as
\begin{equation}
    \hat{\mathcal{V}} = \hat{\mathcal{V}}_0 + \hat{\mathcal{V}}_\text{hol}.
\end{equation}
The first term describes the non-interacting electron potential energy
\begin{equation}
    \hat{\mathcal{V}}_0 = \sum_\sigma \sum_{\mathbf{i},\gamma} (\epsilon_{\mathbf{i},\gamma} - \mu) \hat{n}_{\sigma,\mathbf{i},\gamma},
\end{equation}
where $\hat{n}_{\sigma,\mathbf{i},\gamma} = \hat{c}_{\sigma,\mathbf{i},\gamma}^{\dagger} \hat{c}_{\sigma,\mathbf{i},\gamma}^{\phantom\dagger}$ is the spin-$\sigma$ electron number operator for orbital $\gamma$ in unit cell $\mathbf{i}$, $\epsilon_{\mathbf{i},\gamma}$ is the corresponding site energy, and $\mu$ the chemical potential. The second term describes a potential energy coupling between the electrons and the lattice, such as those that arise in the Holstein or Fr{\"o}hlich models. This term can be defined using one of two distinct and inequivalent parameterizations
\begin{equation}\label{eq:holstein_coupling}
    \hat{\mathcal{V}}_\text{hol} =
    \begin{cases}
        \sum_{\mathbf{i},\eta} \sum_{\mathbf{j},\gamma} \left[
            \sum_{n=1,3} \tilde{\alpha}_{n,(\mathbf{i},\eta),(\mathbf{j},\gamma)} \hat{X}_{\mathbf{i},\eta}^n (\hat{n}_{\mathbf{j},\gamma}-1) + \sum_{n=2,4} \tilde{\alpha}_{n,(\mathbf{i},\eta),(\mathbf{j},\gamma)} \hat{X}_{\mathbf{i},\eta}^n \hat{n}_{\mathbf{j},\gamma}
        \right] \\
        \sum_{\mathbf{i},\eta} \sum_{\mathbf{j},\gamma} \sum_{n=1}^{4} \tilde{\alpha}_{n,(\mathbf{i},\eta),(\mathbf{j},\gamma)} \hat{X}_{\mathbf{i},\eta}^n \hat{n}_{\mathbf{j},\gamma},
    \end{cases}
\end{equation}
where $\hat{n}_{\mathbf{j},\gamma} = (\hat{n}_{\uparrow, \mathbf{j},\gamma} + \hat{n}_{\downarrow, \mathbf{j},\gamma})$ and $\tilde{\alpha}_{n,(\mathbf{i},\eta),(\mathbf{j},\gamma)}$ controls the strength of the \gls*{eph} coupling. The first parameterization in Eq.~\eqref{eq:holstein_coupling} maintains particle-hole symmetry in single-band models such that $\mu = 0$ corresponds to half-filling, while the second does not. 

\section{Quantum Monte Carlo Method}\label{sec:qmc}

This section describes the \gls*{QMC} method implemented in the \gls*{smoqyelphqmc} package in detail. As with other packages released under the SmoQy Suite, here we take an axiomatic approach to describing the established underlying algorithms and refer the reader to the relevant references for complete derivations. Throughout, we define our units such that $\hbar = k_\mathrm{B} = 1$. 

\subsection{A Review of Determinant Quantum Monte Carlo}

We begin by first reviewing how the \gls*{DQMC} method is used to simulate spin-symmetric \gls*{eph} Hamiltonians. 

In \gls*{DQMC} simulations, the partition function is approximated as the path-integral in discretized imaginary-time
\begin{equation}\label{eq:Z_dqmc}
    Z = \tr \left[ e^{-\beta \hat{\mathcal{H}}} \right] = \tr \left[ \left(e^{-\Delta\tau \hat{\mathcal{H}}}\right)^{L_\tau} \right]
    \approx \int\mathcal{D}x \ e^{-\left[S_\text{ph}(x)+S_\text{hol}(x)\right]} |\det(M)|^2 + \mathcal{O}(\Delta\tau^2),
\end{equation}
where $\beta = 1/T = L_\tau \Delta\tau$\revise{, $T$ is the temperature, and $L_\tau$ is the length of the discretized imaginary-time axis}~\cite{Blankenbecler1981montecarlo, White1989numerical, Scalettar1989Competition}. Here, the differential $\mathcal{D}x$ denotes the path-integral over the phonon field configurations $x = \{ x_{0,0}, \dots, x_{l,i}, \dots, x_{L_\tau-1, \mathcal{N}_\text{ph}-1} \}$. These fields are then sampled during a \gls*{DQMC} simulation, with the integrand in Eq.~\eqref{eq:Z_dqmc} serving as the Monte Carlo weight.

The phononic action $S_\text{ph}(x)$ appearing in Eq.~\eqref{eq:Z_dqmc} describes the \revise{non-interacting} lattice degrees of 
freedom. It is given by
\begin{equation}\label{eq:Sph}
    S_\text{ph}(x) = S_\text{qho}(x) + \revise{S_\text{anh}(x)} + S_\text{disp}(x).
\end{equation}
The first term
\begin{equation}\label{eq:S_qho}
    S_\text{qho}(x) = \Delta\tau\sum_{l=0}^{L_\tau-1}\revise{\sum_{i=0}^{\mathcal{N}_\text{ph}-1}}\left[
            \frac{M_i}{2}\left(\frac{x_{l+1,i}-x_{l,i}}{\Delta\tau}\right)^2 + \frac{M_i\Omega_i^2}{2}x_{l,i}^2
        \right]
\end{equation}
is the action arising from the term $\hat{\mathcal{U}}_\text{qho}$ in the Hamiltonian [see Eq.~\eqref{eq:U_qho}]. Here, for convenience, we have adopted a combined index $i = (\mathbf{i},\nu)$ to refer to phonon mode $\nu$ in unit cell $\mathbf{i}$ in the lattice. The second term
\begin{equation}
    S_\text{anh}(x) = \Delta\tau\sum_{l=0}^{L_\tau-1}\revise{\sum_{i=0}^{\mathcal{N}_\text{ph}-1}}\left[
            \frac{M_i\Omega_{\revise{\mathrm{a}},i}^2}{24}x_{l,i}^4
        \right]
\end{equation}
results from the term $\hat{\mathcal{U}}_\text{anh}$ in the Hamiltonian [see Eq.~\eqref{eq:U_anh}]. Finally, the third term contributing to phononic action is
\begin{equation}
    S_\text{disp}(x) = \Delta\tau \sum_{l=0}^{L_\tau-1}\revise{\sum_{i=0}^{\mathcal{N}_\text{ph}-2}\sum_{j=i+1}^{\mathcal{N}_\text{ph}-1}}\left[
            \frac{M_i M_j}{M_i + M_j}\left( \tilde{\Omega}_{i,j}^2(x_{l,i}-\xi_{i,j}x_{l,j})^2 + \frac{1}{12}\tilde{\Omega}_{\revise{\mathrm{a}},i,j}^2(x_{l,i}-\xi_{i,j}x_{l,j})^4 \right)
        \right],
\end{equation}
which comes from the dispersive coupling between phonon modes introduced by the $\hat{\mathcal{U}}_\text{disp}$ term [see Eq.~\eqref{eq:U_disp}]. 

Equation~\eqref{eq:Z_dqmc} also has an additional term
\begin{equation}\label{eq:Shol_1}
    S_\text{hol}(x) = \sum_{l=0}^{L_\tau-1} \sum_{r=0}^{\mathcal{N}-1} S_{\text{hol},r}(x_l),
\end{equation}
with
\begin{equation}\label{eq:Shol_2}
    S_{\text{hol},r}(x_l) = \begin{cases}
        -\Delta\tau \sum_{i=0}^{\mathcal{N}_\text{ph}-1} \left(\tilde{\alpha}_{1,i,r} x_{l,i} +  \tilde{\alpha}_{3,i,r} x_{l,i}^3 \right)\\
        0
    \end{cases}. 
\end{equation}
This term stems from the two possible parameterizations for the \gls*{eph} term $\hat{V}_\text{hol}$ defined in Eq.~\eqref{eq:holstein_coupling}, where $r = (\mathbf{r}, \gamma)$ refers to an orbital $\gamma$ in unit cell $\mathbf{r}$ in the lattice.

The fermionic contribution to the Monte Carlo weight used in \gls*{DQMC} simulations is given by the fermion determinants 
appearing in Eq.~\eqref{eq:Z_dqmc}. In general, there is one matrix and corresponding determinant associated with each fermion spin-species; however, in the case of spin-symmetric \gls*{eph} Hamiltonians, the fermion determinant and the corresponding matrices for each spin species are identical such that $|\det(M_\uparrow)\det(M_\downarrow)| = |\det(M_\uparrow)|^2\equiv |\det(M)|^2$. 
This single $\mathcal{N} \times \mathcal{N}$ matrix is given by
\begin{equation}\label{eq:M_dqmc}
    M = I + B_{L_\tau-1} \dots B_l \dots B_1 B_0,
\end{equation}
where the propagator matrices are given by either
\begin{equation}\label{eq:sym_propagator}
    B_l = [e^{-\Delta\tau K_l/2}]^\dagger e^{-\Delta\tau V_l} e^{-\Delta\tau K_l/2}
\end{equation}
or
\begin{equation}\label{eq:asym_propagator}
    B_l = e^{-\Delta\tau V_l} e^{-\Delta\tau K_l}.
\end{equation}
Here, $V_l = V(x_l)$ and $K_l = K(x_l)$ are the electron potential and kinetic energy matrices, respectively, associated with the $\tau = l \cdot \Delta\tau$ imaginary-time slice. Note that while \gls*{smoqyelphqmc} supports both definitions for the propagator matrices, the remainder of this paper \revise{assumes} that Eq.~\eqref{eq:sym_propagator} is used. For more information \revise{on} details of the algorithm when Eq.~\eqref{eq:asym_propagator} is used, we refer \revise{the reader} to Ref.~\cite{Cohen-Stead2022Fast}.

The electron potential energy matrix $V_l$ is diagonal in orbital space; its matrix elements are
\begin{equation}
    [V_l]_{i,i} = \epsilon_i - \mu + \sum_{p}\sum_{n=1}^{4} \tilde{\alpha}_{n,p,i} x_{l,p}^n
\end{equation}
where $i = (\mathbf{i}, \gamma)$ maps between a matrix index and an orbital $\gamma$ in the unit cell $\mathbf{i}$ and the sum over $p$ runs over all phonon modes in the lattice.

The electron kinetic energy matrix $K_l$ is symmetric with zeros along the diagonal; its matrix elements are 
\begin{equation}\label{eq:hopping matrix}
    [K_{l}]_{i,j} = [K_{l}]_{j,i}  = -t_{l,i,j}
\end{equation}
for $i>j$, where
\begin{equation}\label{eq:t_eff}
    t_{l,i,j} = t_{i,j} - \sum_{\nu_i, \eta_j} \sum_{n=1}^4 \alpha_{n,(i,\nu_i),(j,\eta_j)} (x_{l,\nu_i} - x_{l,\eta_j})^n
\end{equation}
is the total effective hopping between orbitals $i$ and $j$ in the lattice. Here, the sum over $\nu_i \ (\eta_j)$ runs over phonon modes sharing a unit cell with orbital $i \ (j)$ in the lattice.

Various sampling schemes---from fast local updates~\cite{Scalettar1989Competition} to self-learning Monte Carlo~\cite{Chen2018Symmetryenforced, Li2019Accelerating} to Langevin~\cite{Batrouni2019Langevin, Cohen-Stead2020Langevin} and \gls*{HMC}~\cite{Beyl2018Revisiting, Cohen-Stead2024SmoQyDQMCjl} methods---have been used to sample the phonon fields $x$ in \gls*{DQMC} simulations. However, owing to the presence of the fermion determinant in Eq.~\eqref{eq:Z_dqmc}, and the fact that the number of $B_l$ propagator matrices scales linearly with the inverse temperature $\beta \sim L_\tau$, all of these methods unavoidably have a computational cost that scales \revise{as} $\mathcal{O}(\beta \mathcal{N}^3)$.

\subsection{The Checkerboard Approximation}\label{sec:checkerboard}

In practice, \gls*{smoqyelphqmc} does not calculate the exponential of the electron kinetic energy matrices $e^{-\Delta\tau K_l/2}$ exactly. Instead, it approximates them using the sparse \gls*{MSCHK} method, which introduces an additional $\mathcal{O}(\Delta\tau^2)$ approximation~\cite{Lee2013Minimal}. This approximation is done for two reasons. First, when \gls*{SSH}-like \gls*{eph} interactions are included in the model, corresponding to $\hat{\mathcal{K}}_\text{ssh} \ne 0$, updates to the phonon fields would require recomputing the $e^{-\Delta\tau K_l/2}$ matrices at a computational cost that scales \revise{as} $\mathcal{O}(\beta \mathcal{N}^3)$. Second, because $e^{-\Delta\tau K_l/2}$ is a dense matrix, the cost of evaluating a matrix-vector product involving the fermion determinant matrix $M$ in Eq.~\eqref{eq:M_hmc} would scale as $\mathcal{O}(\beta \mathcal{N}^2)$ even in the absence of \gls*{SSH}-like \gls*{eph} interactions ($\hat{\mathcal{K}}_\text{ssh} = 0$). Applying the \gls*{MSCHK} approximation reduces the computational cost of both recomputing the $e^{-\Delta\tau K_l/2}$ matrices and evaluating matrix-vector products involving $M$ to a computational cost that scales linearly in both the inverse temperature and system size $\mathcal{O}(\beta \mathcal{N})$.

To understand the checkerboard approximation, it is first useful to write the kinetic energy matrix in the form
\begin{equation}
    K_l = \sum_b k_{l,b},
\end{equation}
where the sum over $b$ runs over all of the individual hopping terms in the Hamiltonian and
\begin{equation}
    k_{l,b} = \left[\begin{array}{ccccc}
        \ddots & \vdots &  & \vdots\\
        \dots & 0 & \dots & -t_{l,i_b,j_b} & \dots\\
        & \vdots & \ddots & \vdots\\
        \dots & -t_{l,i_b,j_b} & \dots & 0 & \dots\\
        & \vdots &  & \vdots & \ddots
    \end{array}\right]
\end{equation}
is the $\mathcal{N}\times \mathcal{N}$ matrix representation of the corresponding hopping with only two non-zero entries. The exponential of each matrix is given by 
\begin{equation}\label{eq:exp(-dtau k_b)}
    e^{-\Delta\tau k_{l,b}/2} =
    \left[\begin{array}{ccccc}
        1 & \vdots &  & \vdots\\
        \dots & \cosh\left(\frac{\Delta\tau}{2} t_{l,i_b,j_b}\right) & \dots & \sinh\left(\frac{\Delta\tau}{2} t_{l,i_b,j_b}\right) & \dots\\
         & \vdots & 1 & \vdots\\
        \dots & \sinh\left(\frac{\Delta\tau}{2} t_{l,i_b,j_b} \right) & \dots & \cosh\left(\frac{\Delta\tau}{2} t_{l,i_b,j_b}\right) & \dots\\
         & \vdots &  & \vdots & 1
    \end{array}\right].
\end{equation}

Next, the hopping integrals $b$ must be sorted into groups or ``colors,'' such that integrals of a given color do not overlap or touch. This condition corresponds to the situation where the hopping matrices $k_{l,b}$ of the same color commute with each other. The task of constructing these groups can be reduced to the edge coloring problem in graph theory. It is important to use the minimum number of colors, as this improves the accuracy of the checkerboard approximation~\cite{Lee2013Minimal}. However, the precise composition of each color is not unique, and some coloring schemes are better than others~\cite{Beyl2020Hybrid}. In \gls*{smoqyelphqmc}, the colors are assigned by systematically iterating over the unit cells in the lattice, and assigning a color to each bond with a site contained in the current unit cell; for more information, we refer the reader to our \href{https://github.com/cohensbw/Checkerboard.jl.git}{\texttt{Checkerboard.jl}} package.

Having grouped the bonds into colors, the kinetic energy matrix may be conveniently expressed in the form
\begin{equation}
    K_{l} = \sum_{c=1}^{N_c} K_{l,c} = \sum_{c=1}^{N_c} \left[ \sum_{b_c} k_{l,b_c} \right],
\end{equation}
where the sum over $c$ runs over the different bond colors and the sum over $b_c$ runs over hopping integrals of color $c$. As each individual hopping matrix $k_{l,b_c}$ commutes with every other one of the same color, it then follows that 
\begin{equation}
    e^{-\Delta\tau K_{l,c}/2} = \prod_{b_c} e^{-\Delta\tau k_{l,b_c}/2},
\end{equation}
where $e^{-\Delta\tau K_{l,c}/2}$ is a sparse matrix.
Finally, the \gls*{MSCHK} approximation is given by
\begin{equation}
    e^{-\Delta\tau K_l/2} \approx \prod_{c=1}^{N_c} e^{-\Delta\tau K_{l,c}/2} + \mathcal{O}(\Delta\tau^2),
\end{equation}
in which the exponentiated kinetic energy matrix is represented as a product of sparse matrices. At this level, the ordering of the exponentiated sparse color matrices $e^{-\Delta\tau K_{l,c}/2}$ does not matter.

\subsection{Replacing the Fermion Determinant with Pseudofermion Fields}

Evaluating the fermion determinant appearing in Eq.~\eqref{eq:Z_dqmc} is a significant computational bottleneck in standard \gls*{DQMC} simulations. This section discusses a technique for replacing the exact determinant calculation with a complex multivariate Gaussian integral. While this introduces additional pseudofermion field degrees of freedom that will need to be sampled during a Monte Carlo simulation, it also enables \gls*{QMC} simulations that can scale near-linearly with system size.

The first step is to replace the $\mathcal{N} \times \mathcal{N}$ fermion determinant matrix definition in Eq.~\eqref{eq:M_dqmc} by
\begin{equation}\label{eq:M_hmc}
    M=\left(\begin{array}{ccccc}
        I &  &  &  & B_{0}\\
        -B_{1} & I\\
        & -B_{2} & \ddots\\
        &  & \ddots & \ddots\\
        &  &  & -B_{L_{\tau}-1} & I
    \end{array}\right),
\end{equation}
which shares the same determinant but is now an expanded $\mathcal{V}\times \mathcal{V}$ matrix, with $\mathcal{V} = \mathcal{N}\cdot L_\tau$. It is now useful to rewrite Eq.~\eqref{eq:Z_dqmc} in the form
\begin{equation}\label{eq:Z_hmc}
    Z \approx \int\mathcal{D}x \ e^{-S_\text{ph}(x)} |\det(M\Lambda)|^2 + \mathcal{O}(\Delta\tau^2),
\end{equation}
where a new matrix $\Lambda$ is introduced that satisfies the constraint $\revise{(\det \Lambda)^2} = e^{-S_\text{hol}(x)}$, with $S_\text{hol}(x)$ defined in Eq.~\eqref{eq:Shol_1}. Adopting the convention used in Ref.~\cite{Cohen-Stead2022Fast}, we use the sparse matrix
\begin{equation}\label{eq:Lambda}
    \Lambda_{(i,l),(i',l')} = \delta_{l+1,l'}\delta_{i,i'}\left(2 \delta_{l',0} - 1\right) e^{-S_{\text{hol},i'}(x_{l'})/2}
\end{equation}
with inverse
\begin{equation}\label{eq:inv_Lambda}
    \Lambda_{(i,l),(i',l')}^{-1} = \delta_{l,l'+1}\delta_{i,i'}\left(2 \delta_{l,0} - 1\right) e^{+S_{\text{hol},i}(x_{l})/2},
\end{equation}
where the definition for $S_{\text{hol},i}(x_l)$ is given in Eq.~\eqref{eq:Shol_2}.

To proceed, we now need to make use of the fact that the determinant of a $D$-dimensional positive-definite matrix $\Sigma$ may be expressed as the complex multivariate Gaussian integral
\begin{equation}\label{eq:gaussian_integral}
    \det \Sigma = \frac{1}{\pi^D} \int_{\mathbb{C}^{D}} \mathcal{D}\Phi \ e^{-\Phi^\dagger \Sigma^{-1} \Phi}
\end{equation}
over complex-valued vectors $\Phi$. The matrix $\Sigma$ may also be interpreted as defining the covariance matrix for a complex multivariate Gaussian probability distribution
\begin{equation}\label{eq:gaussian_distribution}
    P(\Phi) = \frac{1}{\pi^D \det \Sigma} \ e^{-\Phi^\dagger \Sigma^{-1} \Phi}
\end{equation}
with mean $\langle \Phi \rangle = 0$.

Next, we can replace the fermion determinant appearing in Eq.~\eqref{eq:Z_hmc} by a complex multivariate Gaussian integral using Eq.~\eqref{eq:gaussian_integral} to express the partition function 
as an integral over all possible phonon field configurations $x$ and complex-valued pseudofermion fields $\Phi$\begin{equation}
    Z \approx \int \mathcal{D}\Phi \ \mathcal{D}x \ e^{-S(x,\Phi)}  + \mathcal{O}(\Delta\tau^2). 
\end{equation}
The total effective action in the above expression can be written as the sum of bosonic and fermionic contributions 
\begin{equation}\label{eq:total_action}
    S(x,\Phi) = S_\text{ph}(x) + S_\text{F}(x,\Phi).
\end{equation}
Here, the bosonic action $S_\text{ph}(x)$ is defined in Eq.~\eqref{eq:Sph} and the fermionic action is given by
\begin{equation}\label{eq:fermionic_action}
    S_\text{F}(x,\Phi) = \Phi^\dagger \Sigma^{-1} \Phi^{\phantom \dagger}, 
\end{equation}
where $\Sigma = A^T A$ and $A = M \Lambda$.
Therefore, for fixed phonon field configuration $x$, Eq.~\eqref{eq:gaussian_distribution} tells us that the pseudofermion fields can be directly sampled according to
\begin{equation}\label{eq:sample_phi}
    \Phi = A^T R,
\end{equation}
where $R$ is a complex vector sampled from a multivariate complex Gaussian distribution with a covariance given by the identity matrix.

Evaluation of the bosonic contribution to the total action $S(x,\Phi)$ is straightforward, as $S_\text{ph}(x)$ is a simple scalar function. Evaluating Eq.~\eqref{eq:fermionic_action} to calculate the fermionic contribution $S_\text{F}(x,\Phi)$ is more involved. It is useful to express Eq.~\eqref{eq:fermionic_action} in the form
\begin{equation}\label{eq:Sf_compact}
    \begin{aligned}
        S_\text{F}(x,\Phi) & = \Phi^\dagger \Sigma^{-1} \Phi \\
        & = \Phi^\dagger [A^T A]^{-1} \Phi \\
        & = \Phi^\dagger \Lambda^{-1} [M^T M]^{-1} \Lambda^{-T} \Phi \\
        & = b^\dagger D^{-1} b \\
        & = b^\dagger v, 
    \end{aligned}
\end{equation}
where $D = M^T M$, $b = \Lambda^{-T}\Phi$, and $v = D^{-1} b$. While the vector $b$ can be easily calculated using Eq.~\eqref{eq:inv_Lambda}, calculating $v$ requires solving the linear system
\begin{equation}\label{eq:linear_system}
    D \ v = b
\end{equation}
with an iterative solver like the \gls*{CG} method.
Additionally, to efficiently sample the phonon fields in the \gls*{QMC} simulations, it is necessary to evaluate the derivative of $S_F(x,\Phi)$ with respect to each phonon field $x_{l,i}$. With this in mind, we use the matrix identity \revise{$dC^{-1} = -C^{-1} (dC) C^{-1}$} to express the derivative as
\begin{equation}\label{eq:fermionic_action_derivative}
\begin{aligned}
    \frac{\partial S_F}{\partial x_{l,i}} & = \Phi^\dagger \left[ \frac{\partial \Sigma^{-1}}{\partial x_{l,i}} \right] \Phi = -2 \ \text{Re} \left\{ \left[A\Psi\right]^\dagger \left(\frac{\partial M}{\partial x_{l,i}}\right)  \Lambda \Psi + \left[M^T A\Psi\right]^\dagger \left(\frac{\partial \Lambda}{\partial x_{l,i}}\right) \Psi \right\},
\end{aligned}
\end{equation}
where $\Psi = \Sigma^{-1}\Phi = \Lambda^{-1} v$.

At this point, \revise{it is important to note} that the definition for the enlarged fermion determinant matrix in Eq.~\eqref{eq:M_hmc} is quite sparse. When combined with the sparse checkerboard approximation for the exponentiated kinetic energy matrix of a tight-binding model with finite range hopping integrals, the number of non-zero matrix elements scales linearly with both the system size and inverse temperature. \revise{Therefore, the cost of evaluating a matrix-vector product $M v$ scales as $\mathcal{O}(\beta N)$.} As the \gls*{CG} method requires only matrix-vector products, the required computational cost \revise{per CG iteration} to evaluate the action $S(x,\Phi)$, as well as its derivatives with respect to the phonon fields $x_{l,i}$, \revise{also scales linearly}. It has been observed that the number of required iterations to solve Eq.~\eqref{eq:linear_system} only weakly depends on the system size and temperature for many \gls*{eph} models, particularly in the adiabatic regime relevant to most real materials. Therefore, the final computational complexity \revise{can approach linear scaling} in both the system size and the inverse temperature \revise{as the adiabatic limit is approached}~\cite{Cohen-Stead2022Fast, Batrouni2019Langevin, Beyl2018Revisiting}.

\subsection{Efficient Hybrid Quantum Monte Carlo Updates}

To efficiently sample the phonon fields, the \gls*{smoqyelphqmc} package mainly relies on performing \gls*{HQMC} updates with \gls*{EFA}, which will be referred to as \gls*{EFA-HQMC} updates. These \gls*{EFA-HQMC} updates are performed by evolving a fictitious Hamiltonian dynamics to propose global updates to every phonon field simultaneously. The algorithm for performing these updates is \revise{very similar} to \revise{Algorithm~5} in Ref.~\cite{Cohen-Stead2024SmoQyDQMCjl} with \revise{an} important difference: new pseudofermion fields $\Phi$ are sampled at the  start of the update using Eq.~\eqref{eq:sample_phi}, which are then treated as a constant during the remainder of the update, with the total action used to perform the update then given by Eq.~\eqref{eq:total_action}. As a result, the computationally expensive operation is no longer calculating the inverse of the fermion determinant matrix, but instead solving the linear system in Eq.~\eqref{eq:linear_system} to obtain the fermionic action in Eq.~\eqref{eq:fermionic_action} and the partial derivatives with respect to the phonon fields in Eq.~\eqref{eq:fermionic_action_derivative}. 

Another difference is that because the \gls*{EFA-HQMC} updates in \gls*{smoqyelphqmc} use a modified fermionic action relative to \gls*{smoqydqmc} that now depends on a randomly sampled pseudofermion field $\Phi$, the same time-step $\Delta t$ used in the update may result in a different acceptance rate. More specifically, as $\Phi$ is essentially a random vector, this results in a noisier potential energy landscape, which typically requires using a slightly smaller time-step $\Delta t$ to achieve the same acceptance rate \revise{than would be needed} in a comparable simulation using \gls*{smoqydqmc}. Ultimately, as before, the time-step $\Delta t$ will need to be tuned by the user to achieve a desired acceptance rate.

\subsection{\revise{Reflection, Swap and Radial Updates}}

As with the \gls*{smoqydqmc} package, \gls*{smoqyelphqmc} also provides support for additional \revise{reflection, swap and radial} updates.  As discussed in \revise{Refs.~\cite{Cohen-Stead2024SmoQyDQMCjl,Ostmeyer2025Exponential,Temmen2025Fully}}, these types of global moves can help the simulation tunnel through nodal surfaces and alleviate ergodicity issues that can arise in \gls*{HQMC} simulations of \gls*{eph} models. Note that the accept/reject step for these global moves can be evaluated using the procedures discussed in the previous section and does not modify the \revise{scaling of the \gls*{QMC} method.} 

\section{Kernel Polynomial Method Preconditioner}\label{sec:kpm_preconditioner}
As discussed in Sec.~\ref{sec:qmc}, the most computationally expensive step in the \gls*{QMC} simulations performed using \gls*{smoqyelphqmc} is solving the linear system of the form shown in Eq.~\eqref{eq:linear_system},
\begin{equation}
    D v = b,
\end{equation}
where $D = M^T M$ and  $v$ is an unknown vector. In this section, we introduce a preconditioner $P$ that accelerates this process by allowing us to solve the left-preconditioned system
\begin{equation}
    P^{-1} D v = P^{-1} b
\end{equation}
using the preconditioned \gls*{CG} method. This approach accelerates the calculation when the matrix $P^{-1} D$ is better conditioned than $D$ and matrix-vector products of the form $P^{-1} x$ can be efficiently evaluated.

\subsection{An Adiabatic Limit Inspired Preconditioner}

Our preconditioner is inspired by the observation that the \gls*{QHO} action $S_\text{qho}(x)$, defined in Eq.~\eqref{eq:S_qho}, suppresses fluctuations in imaginary-time by coupling together phonon fields at adjacent imaginary-time slices. \revise{This is especially true in the adiabatic limit, which is obtained by sending the phonon mass to infinity $(M\rightarrow \infty)$ and the phonon frequency to zero $(\Omega \rightarrow 0)$, while keeping the corresponding spring constant fixed $(k = M\Omega^2 = \text{const.})$.} In this limit the imaginary-time fluctuations are entirely suppressed, and the corresponding \gls*{eph} Hamiltonian permits a semiclassical description.
Functionally, approaching the adiabatic limit results in $D = M^T M$ and $M$ approaching the semiclassical matrices $\mathcal{D} = \mathcal{M}^T\mathcal{M}$ and
\begin{equation}\label{eq:M_bar}
    \mathcal{M} =\left(\begin{array}{ccccc}
        I &  &  &  & \mathcal{B}_0\\
        -\mathcal{B}_0 & I\\
        & -\mathcal{B}_0 & \ddots\\
        &  & \ddots & \ddots\\
        &  &  & -\mathcal{B}_0 & I
    \end{array}\right)
\end{equation}
respectively, where the definition
\begin{equation}\label{eq:B0}
    \mathcal{B}_0 = \frac{1}{L_\tau} \sum_{l=0}^{L_\tau - 1} B_l
\end{equation}
effectively averages out the fluctuations in imaginary-time. (For the remainder of this section, matrices written in a calligraphic font, e.g., $\mathcal{D}$ and $\mathcal{M}$, will denote quantities with their imaginary-time fluctuations suppressed.)

Our preconditioner systematically approximates the adiabatic limit $P \approx \mathcal{D}$ by applying the \gls*{KPM} in conjunction with a unitary transformation
\begin{equation}\label{eq:unitary_transform}
    Q_{n,l} = \frac{1}{\sqrt{L_\tau}}e^{-\mathrm{i}\frac{\pi(2n+1)l}{L_\tau} } = \frac{1}{\sqrt{L_\tau}}e^{-\mathrm{i} \omega_n \tau_l }
\end{equation}
that block diagonalizes the matrices $\mathcal{D}$ and $\mathcal{M}$, where $\tau_l = \Delta\tau\cdot l$ and $\omega_n = (2n+1)\pi/\beta$.
Moving forward, matrices transformed to this unitary basis will be denoted by a tilde,  i.e. $\tilde{A} = Q \ A \ Q^\dagger$. It is useful to express this transformation as the product of two other unitary transformations, $Q = F\Theta$. The first unitary factor is the diagonal matrix
\begin{equation}
    \Theta_{l,l'} = \delta_{l,l'} e^{-\mathrm{i}\pi l/L_\tau},
\end{equation}
where $\Theta M \Theta^\dagger$ has the same block structure as $M$ but with a uniform $-e^{-\mathrm{i}\pi/L_\tau}$ factor appearing in front of each $B_l$ block matrix. The second unitary factor is the imaginary-time Fourier transform matrix
\begin{equation}
    F_{n,l} = \frac{1}{\sqrt{L_\tau}}e^{-\mathrm{i}\frac{2\pi}{L_\tau}nl}.
\end{equation}
Importantly, since the matrix $\Theta$ is a diagonal and $F$ can be efficiently applied to a vector via a \gls*{FFT}, the computational cost to evaluate a matrix-vector product $Q v$ scales as $\mathcal{O}(\mathcal{N}L_\tau \log (L_\tau))$.

The $\mathcal{N} \times \mathcal{N}$ blocks of the transformed fermion determinant matrix $\tilde{M} = Q M Q^\dagger$ are given by
\begin{equation}
    \tilde{M}_{n,m} = \delta_{n,m}I - e^{-{\rm i}\omega_{m}\Delta\tau} \mathcal{B}_{n-m},
\end{equation}
while the blocks of $\tilde{D} = Q D Q^\dagger$ are
\begin{equation}
    \tilde{D}_{n,m} = \delta_{n,m}I-e^{-i\Delta\tau\omega_{\revise{m}}}\mathcal{B}_{n-m}^{\phantom{\dagger}}-e^{+\mathrm{i}\Delta\tau\omega_{\revise{n}}}\mathcal{B}_{\revise{m-n}}^{\dagger}+\revise{e^{i\Delta\tau(\omega_n-\omega_m)}}\sum_{n'=0}^{L_\tau-1}\mathcal{B}_{\revise{n'-n}}^{\dagger}\mathcal{B}_{n'-m}^{\phantom{\dagger}}.
\end{equation}
In the two equations above
\begin{equation}
    \mathcal{B}_n = \frac{1}{\sqrt{L_\tau}}\sum_{l=0}^{L_\tau-1}F_{n,l} B_l,
\end{equation}
which is consistent with the definition for $\mathcal{B}_0$ in Eq.~\eqref{eq:B0}.
It \revise{is} also straightforward to show that $\tilde{\mathcal{M}} = Q\mathcal{M}Q^\dagger$ is block diagonal such that
\begin{equation}
    \tilde{\mathcal{M}}_{n} = \revise{\tilde{M}_{n,n} = }\left( I - e^{-{\rm i}\Delta\tau \omega_{n}} \mathcal{B}_0 \right).
\end{equation}
In the $Q$ basis, \revise{we then define the preconditioner to represent the} block diagonal matrix
\begin{equation}
     \tilde{\mathcal{D}}_{n} = \revise{\tilde{\mathcal{M}}_n^\dagger \tilde{\mathcal{M}}_n^{\phantom\dagger} =  I - 2\mathcal{B}_0\cos(\Delta\tau\omega_n) + \mathcal{B}_0^2},
\end{equation}
\revise{which approximates the diagonal blocks of $\tilde{D}$ with a difference
\begin{equation}
    \tilde{D}_{n,n} - \tilde{\mathcal{D}}_{n} = \frac{1}{L_\tau}\sum_l B_l^2 - \mathcal{B}_0^2 = \frac{1}{L_\tau}\sum_l B_l^2 - \left( \frac{1}{L_\tau}\sum_l B_l \right)^2
\end{equation}
that goes to zero in the adiabatic limit.}

Functionally, this means that the preconditioner $P = Q^\dagger\tilde{P}Q$ needs to be block-diagonal in the $Q$ basis $(\tilde{P}_{n,n'} = \delta_{n,n'} \tilde{P}_n)$, with the inverted diagonal blocks approximating
\begin{equation}
    \tilde{P}_n^{-1} \approx \revise{\tilde{\mathcal{D}}_n^{-1}} =  f_n(\mathcal{B}_0),
\end{equation}
where
\begin{equation}\label{eq:fn}
    f_n(b) = \left[1 - 2b\cos(\Delta\tau \omega_n) + b^2\right]^{-1}.
\end{equation}
We also need to be able to efficiently evaluate matrix-vector products of the form $P^{-1}v$.
Section~\ref{sec:kpm_approx} describes how both of these criteria are met by using the \gls*{KPM} to systematically formulate the approximation.

\subsection{Applying the Kernel Polynomial Method}\label{sec:kpm_approx}

The \gls*{KPM} is used to systematically approximate matrix-vector products of the form $f(A) u$ without needing to explicitly construct the matrix $f(A)$, where $f(A)$ is a function of a Hermitian matrix $A$ with a bounded spectrum and $u$ is a vector. In this section, we describe how the \gls*{KPM} method is used to evaluate the matrix-vector products
\begin{equation}\label{eq:matrix_function_vector_product}
    \revise{\tilde{\mathcal{D}}_{n}^{-1}} v = f_n({\mathcal{B}_0}) v,
\end{equation}
where the function $f_{n}(\bullet)$ is defined in Eq.~\eqref{eq:fn}. This approach relies on the fact that $\mathcal{B}_0$, as defined in Eq.~\eqref{eq:B0}, is Hermitian with eigenvalues     $b_{\min} \le b \le b_{\max}$ bounded near 1,
otherwise $\Delta\tau$ would not be small enough for the \revise{Suzuki-Trotter} approximation to be a good approximation~\cite{Cohen-Stead2022Fast}.

We proceed by defining a rescaled matrix
\begin{equation}\label{eq:A0}
    \mathcal{A}_0 = (\mathcal{B}_0 - \bar{b}) / \Delta b,
\end{equation}
with corresponding eigenvalues $a = (b - \bar{b})/\Delta b$ bounded between $-1 \le a \le 1$, where $\Delta b = (b_{\max} - b_{\min})/2$ and $\bar{b} = (b_{\max} + b_{\min})/2$. The next step is to approximate the function
\begin{equation}
    \bar{f}_n(a) = f_n(\Delta b \ a + \bar{b})
\end{equation}
as a Chebyshev polynomial expansion
\begin{equation}\label{eq:fbar_expansion}
    \bar{f}_n(a) \approx \sum_{p=0}^{N_p-1} c_p T_p(a)
\end{equation}
of order $N_p$, where $T_p(x) = \cos(p \arccos x)$ are the Chebyshev polynomials defined on the interval $-1 \le x \le 1$.

The Chebyshev polynomials satisfy the orthogonality relation
\begin{equation}
    \int_{-1}^{+1} w(x) T_{p}(x) T_{p'}(x) dx = q_p \delta_{p,p'},
\end{equation}
where \revise{$w(x) = 1/\sqrt{1-x^2}$} and $q_p = \frac{\pi}{2}(1+\delta_{p,0})$.
Therefore, the expansion coefficients appearing in Eq.~\eqref{eq:fbar_expansion} are given by
\begin{equation}
    c_p = \frac{1}{q_p} \int_{-1}^{+1}  w(x) T_{p}(x) \bar{f}_n(x) dx.
\end{equation}
The coefficients $c_p$ can be evaluated efficiently and accurately via Gauss-Chebyshev quadrature~\cite{Weisse2006Kernel}
\begin{equation}\label{eq:gauss-cheby_cp}
    c_p \approx \frac{\pi}{q_p N_q} \sum_{q=0}^{N_q-1} \cos(p \theta_q) \bar{f}_n(\cos \theta_q),
\end{equation}
using $N_q = 2N_p$ quadrature points $\theta_q = \pi (q + \frac{1}{2})/N_q$. As Eq.~\eqref{eq:gauss-cheby_cp} is equivalent to performing a discrete cosine transformation of the second kind (DCT-II) on the data points $\bar{f}_n(\cos \theta_q)$, all of the coefficients \revise{$c_p$} can be evaluated simultaneously at cost $\mathcal{O}(N_q \log N_q)$ using a package like FFTW~\cite{frigo_1998_fftw}.

 Having constructed the Chebyshev polynomial expansion for $\bar{f}_n(a)$ in Eq.~\eqref{eq:fbar_expansion}, the corresponding representation for the matrix $\revise{\tilde{\mathcal{D}}_n^{-1}} =  f_n(\mathcal{B}_0) = \bar{f}_n(\mathcal{A}_0)$ is simply
\begin{equation}
     \bar{f}_n(\mathcal{A}_0) \approx \revise{\sum_{p=0}^{N_p-1}} c_p T_p(\mathcal{A}_0).
 \end{equation}
A matrix-vector product using this representation may be expressed as
\begin{equation}
    \revise{\tilde{\mathcal{D}}_n^{-1}} u =  \bar{f}_n(\mathcal{A}_0)u \approx \revise{\sum_{p=0}^{N_p-1}} c_p \alpha_p,
\end{equation}
where
\begin{equation}
    \alpha_p = T_p(\mathcal{A}_0) u.
\end{equation}
The three-term Chebyshev polynomial recurrence
\begin{equation}
    T_{p+1}(\mathcal{A}_0) = 2 \mathcal{A}_0 T_{p}(\mathcal{A}_0) - T_{p-1}(\mathcal{A}_0)
\end{equation}
may be used to sequentially generate the $\alpha_p$ vectors,
\begin{equation}
    \alpha_{p+1} = 2 \mathcal{A}_0 \alpha_p - \alpha_{p-1},
\end{equation}
starting from $\alpha_1 = \mathcal{A}_0 u$ and $\alpha_0 = u$.

Having described how the \gls*{KPM} is used to formulate an efficient preconditioning method, a few practical considerations need to be discussed. First, we use the heuristic
\begin{equation}\label{eq:N_p}
    N_p = \revise{\max \left( 1 \ , \ \left\lfloor 2 \Delta b \left( \frac{1+\phi_n}{\phi_n} \right)\right\rfloor\right)}
\end{equation}
to determine the expansion order for $\bar{f}_n(\bullet)$, where $\phi_n = \min(\omega_n\Delta\tau \ , \ 2\pi-\omega_n\Delta\tau)$ and $\left\lfloor \bullet \right\rfloor$ denotes the floor operation. Note that the polynomial order scales linearly with $2\Delta b = (b_{\max} - b_{\min})$ and decays rapidly with $n$, so that the \textit{typical} value of $N_p$ is of order one.

Second, to apply the \gls*{KPM}, the eigen-spectrum bounds $b_{\min}$ and $b_{\max}$ for $\mathcal{B}_0$ need to be specified, which are not known \textit{a priori}. Reasonable approximate bounds can, however, be proposed using the Lanczos method. 
More specifically, whenever the phonon configuration is changed, the Lanczos method is used to re-calculate the minimum $\epsilon_{\min}$ and maximum $\epsilon_{\max}$ eigenvalues of $\mathcal{B}_0$ using $\sim 20$ Lanczos iterations. Then, new eigenvalue bounds
\begin{equation}
    b_{\min}^\prime = (1-\delta)\epsilon_{\min} \quad \text{ and } \quad b_{\max}^\prime = (1+\delta)\epsilon_{\max}
\end{equation}
are calculated using a small buffer $\delta \approx 0.05$. If either
\begin{equation}
    \left| \frac{b_{\min}^\prime-b_{\min}^{\phantom\prime}}{b_{\min}} \right| > \frac{\delta}{2} \quad \text{ or } \quad \left| \frac{b_{\max}^\prime-b_{\max}^{\phantom\prime}}{b_{\max}} \right| > \frac{\delta}{2}
\end{equation}
is true, then the new bounds are used to update the expansion order and coefficients using Eq.~\eqref{eq:N_p} and Eq.~\eqref{eq:gauss-cheby_cp}, respectively.

Future work might consider replacing \gls*{KPM} with the Lanczos method to approximate the matrix-vector product in Eq.~\eqref{eq:matrix_function_vector_product} using techniques similar to those described in Ref.~\cite{Chen2024Lanczos}.

\subsection{Approximating the Imaginary-Time Averaged Propagator Matrix}

The exact imaginary-time averaged propagator matrix $\mathcal{B}_0$, as defined in Eq.~\eqref{eq:B0}, is impractical to construct exactly as it is not representable in a form consistent with the \gls*{MSCHK} approximation introduced in Sec.~\ref{sec:checkerboard}. Therefore, we instead approximate it as 
\begin{equation}\label{eq:B0_approx}
    \mathcal{B}_0 \approx \Gamma^\dagger \Lambda \Gamma + \mathcal{O}(\Delta\tau^2),
\end{equation}
which is consistent with the \gls*{MSCHK} approximation while also remaining Hermitian.
Here,
\begin{equation}
    \Lambda = \frac{1}{L_\tau}\sum_{l=0}^{L_\tau-1} e^{-\Delta\tau V_l}
\end{equation}
is the imaginary-time averaged exponentiated potential energy matrix, where each $e^{-\Delta\tau V_l}$ is diagonal. The $\Gamma$ matrix in Eq.~\eqref{eq:B0_approx}, on the other hand, is a \gls*{MSCHK} approximation for the imaginary-time averaged exponentiated kinetic energy matrix. It is given by
\begin{equation}
    \Gamma = \prod_{c=1}^{N_c} \Gamma_c
\end{equation}
with
\begin{equation}
    \Gamma_c = \prod_{b_c} \left[ \frac{1}{L_\tau} \sum_{l=0}^{L_\tau-1} e^{-\Delta\tau k_{l,b_c}/2} \right],
\end{equation}
where $e^{-\Delta\tau k_{l,b_c}/2}$ is defined in Eq.~\eqref{eq:exp(-dtau k_b)}. The index $c \in [1, N_c]$ refers to the different bond colors used in the \gls*{MSCHK} approximation, as discussed in Sec.~\ref{sec:checkerboard}.

\section{Stochastic Measurements}\label{sec:stoch_meas}

In standard \gls*{DQMC} calculations, electronic correlation measurements are made using the single-particle electron Green's function, whose expectation values for a fixed phonon configuration are given by the matrix elements of the inverted fermion determinant matrix, $G = M^{-1}$. However, in \gls*{smoqyelphqmc} we would like to avoid explicitly constructing this matrix as the computational cost of doing so scales cubically with the matrix dimension. In this section, we discuss an alternative approach in which unbiased stochastic estimates of the matrix elements are computed at a cost that scales nearly linearly with the matrix dimension.

\subsection{The Single-Particle Electron Green's Function}

The single-particle imaginary-time electron Green's function is given by
\begin{equation}\label{eq:greens}
    G_{\sigma,i,j}(\tau,\tau') = \big\langle \hat{\mathcal{T}} \hat{c}_{\sigma, \mathbf{i}, \nu}^{\phantom\dagger}(\tau) \hat{c}_{\sigma, \mathbf{j}, \gamma}^{\dagger}(\tau') \big\rangle =
    \begin{cases}
        {\phantom -}\big\langle \hat{c}_{\sigma, \mathbf{i}, \nu}^{\phantom\dagger}(\tau) \hat{c}_{\sigma, \mathbf{j}, \gamma}^{\dagger}(\tau') \big\rangle \quad \text{if} \quad (\tau - \tau') \ge 0 \\
        - \big\langle \hat{c}_{\sigma, \mathbf{j}, \gamma}^{\dagger}(\tau') \hat{c}_{\sigma, \mathbf{i}, \nu}^{\phantom\dagger}(\tau) \big\rangle \quad \text{if} \quad (\tau - \tau') < 0
    \end{cases}, 
\end{equation}
where $0 \le (\tau, \tau') < \beta$, $\hat{\mathcal{T}}$ is the imaginary-time ordering operator, and $i = (\mathbf{i},\nu)$ and \revise{$j = (\mathbf{j}, \gamma)$} specify the orbitals in the lattice. It is then natural to define a $\mathcal{N} \times \mathcal{N}$ matrix $G_{\sigma}(\tau,\tau')$ with matrix elements given by Eq.~\eqref{eq:greens}. Given these definitions, the single-particle electron Green's functions for a fixed phonon configuration are given by the matrix elements of the inverse fermion determinant matrix $G_\sigma = M^{-1}$, as defined in Eq.~\eqref{eq:M_hmc}, such that
\begin{equation}
    G_\sigma = \footnotesize \begin{bmatrix}
        G_\sigma(0,0) & G_\sigma(0,\Delta\tau) & \dots & G_\sigma(0, \tau_l) & \dots & G_\sigma(0, \beta - \Delta\tau) \\
        G_\sigma(\Delta\tau, 0) & G_\sigma(\Delta\tau, \Delta\tau) & \dots & G_\sigma(\Delta\tau,\tau_l) & \dots & G_\sigma(\Delta\tau, \beta-\Delta\tau)\\
        \vdots & \vdots & \ddots & \vdots & \ddots & \vdots \\
        G_\sigma(\tau_l, 0) & G_\sigma(\tau_l, \Delta\tau) & \dots & G_\sigma(\tau_l, \tau_l) & \dots & \revise{G_\sigma}(\tau_l, \beta-\Delta\tau) \\
        \vdots & \vdots & \ddots & \vdots & \ddots & \vdots \\
        G_\sigma(\beta-\Delta\tau, 0) & G_\sigma(\beta-\Delta\tau, \Delta\tau)  & \dots & G_\sigma(\beta-\Delta\tau, \tau_l) & \dots & G_\sigma(\beta-\Delta\tau, \beta-\Delta\tau)
    \end{bmatrix}\normalsize,
\end{equation}
where $\tau_l = \Delta\tau \cdot l$ for $l \in [0, L_\tau)$.

\subsection{Stochastic Approximation of Matrix Elements}\label{sec:stochastic_vectors}

Given a random vector $\xi$ satisfying $\langle \xi \xi^\dagger \rangle = I$, or equivalently $\langle \xi_i \rangle = 0$ and $\langle \xi_i^{\phantom *} \xi_j^{\revise{*}} \rangle = \delta_{i,j}$, it follows that
\begin{equation}
    G = G \langle \xi \xi^\dagger \rangle = \langle (G \xi) \xi^\dagger \rangle
\end{equation}
for a fixed matrix $G$. Consequently, one can construct unbiased stochastic estimators for the individual matrix elements of $G$ as
\begin{equation}\label{eq:matrix_element_estimation}
    G_{i,j} \approx (G\xi)^{\phantom*}_i \xi^*_j,
\end{equation}
where $\xi^*$ denotes the complex conjugation of each element of the random vector $\xi$ without transposition. Likewise, unbiased stochastic estimates of products of matrix elements can be obtained using pairs of independent random vectors $\xi$ and $\zeta$ as
\begin{equation}\label{eq:matrix_prod_est}
    G_{i,j} G_{k,g} \approx (G\xi)^{\phantom*}_i \xi^*_j (G\zeta)^{\phantom*}_k \zeta^*_g.
\end{equation}
A common choice for sampling each component of the random vectors is to draw them from a standard normal distribution, $\xi_j \sim N(0,1)$. However, in \gls*{smoqyelphqmc} a random phase $\xi_j = e^{{\rm i}\theta_j}$ is used instead, with the angle drawn from $\theta_j \sim \text{Uniform}(0,2\pi)$. Theoretical work has shown that this choice of random vector reduces fluctuations in the stochastic estimates by eliminating variations in the magnitude of the random vectors, as it guarantees $|\xi_i|^2 = 1$~\cite{Iitaka2004Random,Wang2018Gradientbased}.

\subsection{FFT-Accelerated Averages Over Translational Symmetry}\label{sec:fft_avg}

As we will see in the next section, when making correlation measurements, we often want to average over translation symmetry in both space and imaginary time. This usually requires evaluating sums of the form
\begin{equation}\label{eq:translation_avg}
    C_{\boldsymbol{\Delta}} = \frac{1}{V} \sum_{\boldsymbol{i}} a_{\boldsymbol{i}+\boldsymbol{\Delta}} b_{\boldsymbol{i}}
\end{equation}
for all possible $\boldsymbol{\Delta}$, where $a$ and $b$ are complex-valued $(D+1)$-dimensional tensors of size $V = N \cdot L_\tau = (L_1 \cdot \ldots \cdot L_D) \cdot L_\tau$.
Consider this sum within the context of a $D+1$ dimensional periodic Bravais lattice, where the additional dimension corresponds to a discretized imaginary-time axis of length $L_\tau$. Assuming just a single site per unit cell, the location of each site in the lattice may be labeled by integers $0 \le n_d < L_d$, where $L_d$ is the linear system size for physical dimension $d$.
Given this context, the index $\boldsymbol{i} = (n_1, \dots, n_D, l)$ appearing in Eq.~\eqref{eq:translation_avg} also specifies a space-time location, with $0 \le l < L_\tau$ labeling the imaginary-time slice, and $\boldsymbol{\Delta}$ is a space-time displacement. This sum can be expressed as an atypically defined complex circular cross-correlation
\begin{equation}\label{eq:cross_correlation}
    C_{\boldsymbol{\Delta}} = [a \star b]_{\boldsymbol{\Delta}}
\end{equation}
and may be efficiently evaluated using \gls*{FFT}s as
\begin{equation}\label{eq:fft_cross_correlation}
    C_{\boldsymbol{\Delta}} = \frac{1}{\sqrt{V}} \ \mathcal{F}^{-1} \left\{ \mathcal{F}\left\{ a \right\} \odot \mathcal{F}\left\{ b^* \right\}^* \right\}
\end{equation}
where $\mathcal{F}$ is a $(D+1)$-dimensional \gls*{FFT} and $\odot$ denotes the element-wise multiplication of tensors of the same size. This method of evaluating the sum in Eq.~\eqref{eq:translation_avg} for all values of $\boldsymbol{\Delta}$ reduces the computational cost from $\mathcal{O}(V^2)$ to $\mathcal{O}(V \log V )$ instead.

\subsection{Translationally Averaged Correlation Measurements}

In this section we will discuss how \gls*{smoqyelphqmc} efficiently measures electronic \revise{correlation functions}. We will begin by discussing how the single-particle electron Green's function needs to be measured.
However, to do so it is necessary to first account for the Green's function being anti-periodic in imaginary-time, $G_\sigma(\tau+\beta,\tau') = -G_\sigma(\tau,\tau')$. Therefore, it is useful to define an extended matrix
\begin{equation}
    G^\text{ext.}_{\sigma} = \begin{bmatrix}
        {\phantom - }G_\sigma & -G_\sigma \\
        -G_\sigma & {\phantom - }G_\sigma
    \end{bmatrix} = W G_\sigma W^T,
\end{equation}
where
\begin{equation}
    W = \begin{bmatrix}
        {\phantom - }I \\
        -I
    \end{bmatrix}.
\end{equation}
The extended matrix $G^{\text{ext.}}$ effectively doubles the range of the imaginary-time index, $0 \le l < 2L_\tau$, such that both the space and imaginary-time indices become periodic.

In this way, the averaged Green's function that we want to measure is
\begin{equation}\label{eq:greens_avg}
    \mathcal{G}_{\sigma, \boldsymbol{\Delta}} = \frac{1}{2 V} \sum_{\boldsymbol{i}} G^\text{ext.}_{\sigma, \boldsymbol{i} + \boldsymbol{\Delta}, \boldsymbol{i}}
\end{equation}
for all $\boldsymbol{\Delta}$, where $\boldsymbol{i} = (l, i)$ is an index that also specifies a position in both space and imaginary-time and $\boldsymbol{\Delta}$ is a displacement in both space and imaginary-time. Eq.~\eqref{eq:greens_avg} can be estimated using the stochastic method from Sec.~\ref{sec:stochastic_vectors} as
\begin{equation}
    \begin{aligned}
        \mathcal{G}_{\sigma,\boldsymbol{\Delta}} & \approx \frac{1}{2V} \sum_{\boldsymbol{i}} \ [W G_\sigma \xi \xi^\dagger W^T]_{\boldsymbol{i}+\boldsymbol{\Delta},\boldsymbol{i}} \\
        & = \frac{1}{2V} \sum_{\boldsymbol{i}} \ [WG_\sigma \xi]_{\boldsymbol{i}+\boldsymbol{\Delta}} \  [W\xi^*]_{\boldsymbol{i}} \\
        & = \big[ (WG_\sigma \xi) \star (W\xi^*) \big]_{\mathbf{\Delta}},
    \end{aligned}
\end{equation}
which involves the extended vectors
\begin{equation}
    WG_\sigma \xi = \left[ {{\phantom - }G_\sigma\xi \atop -G_\sigma \xi} \right] \quad \text{and} \quad W\xi^{*} = \left[ {{\phantom -}\xi^* \atop -\xi^{*}} \right],
\end{equation}
and can be efficiently evaluated for all displacements $\mathbf{\Delta}$ using Eq.~\eqref{eq:fft_cross_correlation}.

Other types of electronic multi-point correlation functions can be measured by applying Wick's theorem to represent them as sums of products of pairs of single-particle electron Green's functions. This results in three types of sums occurring, each of which can be estimated using Eq.~\eqref{eq:matrix_prod_est}. The first type is
\begin{equation}\label{eq:GR0_GR0}
    \begin{aligned}
        \frac{1}{V}\sum_{\boldsymbol{i}} G_{\sigma,\boldsymbol{i}+\boldsymbol{\Delta}, \boldsymbol{i}} G_{\sigma',\boldsymbol{i}+\boldsymbol{\Delta}, \boldsymbol{i}} & \approx \frac{1}{V} \sum_{\boldsymbol{i}} (G_{\sigma} \xi \xi^\dagger)_{\boldsymbol{i}+\boldsymbol{\Delta}, \boldsymbol{i}} (G_{\sigma'} \zeta \zeta^\dagger)_{\boldsymbol{i}+\boldsymbol{\Delta}, \boldsymbol{i}} \\
        & = \frac{1}{V} \sum_{\boldsymbol{i}} \big[ \revise{(G_\sigma \xi)_{\boldsymbol{i}+\boldsymbol{\Delta}} (G_{\sigma'} \zeta)_{\boldsymbol{i}+\boldsymbol{\Delta}}} \big] \big[ \xi^{*}_{\boldsymbol{i}} \zeta^{*}_{\boldsymbol{i}} \big] \\
        & = \big[ (G_\sigma \xi \odot G_{\sigma'} \zeta) \star (\xi^* \odot \zeta^* ) \big]_{\boldsymbol{\Delta}},
    \end{aligned}
\end{equation}
which can be efficiently evaluated for all displacements $\Delta$ using Eq.~\eqref{eq:fft_cross_correlation} once more. The remaining two types of sums may be estimated as
\begin{equation}
    \begin{aligned}
        \frac{1}{V}\sum_{\boldsymbol{i}} G_{\sigma,\boldsymbol{i}+\boldsymbol{\Delta}, \boldsymbol{i}+\boldsymbol{\Delta}} G_{\sigma',\boldsymbol{i}, \boldsymbol{i}} & \approx \big[ (G_\sigma \xi \odot \xi^*) \star (G_{\sigma'}\zeta \odot \zeta^* ) \big]_{\boldsymbol{\Delta}} \quad \text{and}\\
        \frac{1}{V}\sum_{\boldsymbol{i}} G_{\sigma,\boldsymbol{i}, \boldsymbol{i}+\boldsymbol{\Delta}} G_{\sigma',\boldsymbol{i}+\boldsymbol{\Delta}, \boldsymbol{i}} & \approx \big[ \revise{(\xi^* \odot  G_{\sigma'}\zeta)  \star (G_\sigma \xi \odot \zeta^*)} \big]_{\boldsymbol{\Delta}}.
    \end{aligned}
\end{equation}

\subsection{Reducing Stochastic Errors in Correlation Function Estimates}\label{sec:reducing_stoch_errors}

The stochastic error in Eq.~\eqref{eq:matrix_element_estimation} can be reduced by averaging over $N_\text{rv}$ independent random vectors $[\xi_1, \dots, \xi_{N_\text{rv}}]$,
\begin{equation}
    G_{i,j} \approx \frac{1}{N_\text{rv}} \sum_{n=1}^{N_\text{rv}} \ \revise{(G \xi_n^{\phantom *})_i (\xi_n^*)_{j}}
\end{equation}
where the stochastic error then falls off as $1/\sqrt{N_\text{rv}}$. Similarly, the error in Eq.~\eqref{eq:matrix_prod_est} can be reduced by averaging over all possible pairs of random vectors\revise{, such that $G_{i,j} G_{k,g} \approx  P$ with}
\begin{equation}\label{eq:GR0_GR0_avg}
    \revise{P=} \ \binom{N_\text{rv}}{2}^{-1} \sum_{n<m} (G\xi_n^{\phantom*})^{\phantom*}_{i} (\xi^{*}_{n})^{\phantom*}_j (G\xi_m^{\phantom*})^{\phantom*}_{k} (\xi^*_{m})^{\phantom*}_g,
\end{equation}
where $\binom{N_\text{rv}}{2} = N_\text{rv}(N_\text{rv}-1)/2$.
\revise{Each term in the sum is an unbiased estimate of $G_{i,j} G_{k,g}$ because $\xi_n$ and $\xi_m$ are independent for $n \ne m$.

Although $P$ averages over $\binom{N_\text{rv}}{2} \sim N_\text{rv}^2$ pairs, these estimates are not all independent, as pairs that share a random vector are correlated. As a result, it is not guaranteed that the \gls*{RMS} error of $P$ scales as $N_\text{rv}^{-1}$. To understand when this $N_\text{rv}^{-1}$ scaling can still arise, write the two single-vector estimates as $A_n=\mu_A+a_n$ and $B_n=\mu_B+b_n$, where $\mu_A=G_{i,j}$ and $\mu_B=G_{k,g}$ are the desired matrix elements and $a_n$ and $b_n$ are zero-mean stochastic errors. Using these definitions, the variance of $P$ over realizations of random vectors for a fixed matrix $G$ can be expressed as
\begin{equation}\label{eq:var_P}
    \text{Var}(P) = \mathbb{E}[|P-\mu_A\mu_B|^2] = \langle |Q|^2 \rangle + \langle |L|^2 \rangle,
\end{equation}
where
\begin{equation}\label{eq:Q_and_P}
    \begin{aligned}
        \langle|L|^2\rangle & =\frac{4N_{\rm rv}-2}{3N_{\rm rv}(N_{\rm rv}-1)}
        \left[\frac{1}{\mathcal{N}_A}+\frac{1}{\mathcal{N}_B}
        +\frac{4(N_{\rm rv}-2)}{4N_{\rm rv}-2}\,
        \frac{\operatorname{Re}\rho}{\sqrt{\mathcal{N}_A\mathcal{N}_B}}\right]
        \sigma_A^2\sigma_B^2 \quad \text{and}\\
        \langle |Q|^2 \rangle & = \frac{2}{N_\text{rv}(N_\text{rv}-1)} \ \sigma_A^2\sigma_B^2.
    \end{aligned}
\end{equation}
Here, $\sigma_A^2= \text{Var}(A_n) = \langle|a_n|^2\rangle$ and $\sigma_B^2 = \text{Var}(B_n) = \langle|b_n|^2\rangle$ are the variance of the single-vector estimates $A_n$ and $B_n$, while
\begin{equation}
    \mathcal{N}_A
  = \frac{\sigma_A^2}{|\mu_A|^2}
  = \frac{\sum_{a\neq j}|G_{i,a}|^2}{|G_{i,j}|^2},
\qquad
\mathcal{N}_B
  = \frac{\sigma_B^2}{|\mu_B|^2}
  = \frac{\sum_{b\neq g}|G_{k,b}|^2}{|G_{k,g}|^2}
\end{equation}
are the squared noise-to-signal ratio of these estimates for unimodular noise. Finally,
\begin{equation}
    \rho=\frac{\mu_B\mu_A^*\langle a_n b_n^*\rangle}{|\mu_A\mu_B|\,\sigma_A\sigma_B}
\end{equation}
measures the correlation between the two fluctuations generated by the same
random vector, with $|\rho|\le 1$.

The scaling behavior of Eq.~\eqref{eq:Q_and_P} is
\begin{equation}
    \langle|L|^2\rangle \sim \frac{\sigma_A^2\sigma_B^2}{\mathcal{N}_\text{eff}N_\text{rv}} \quad \text{and} \quad \langle | Q|^2 \rangle \sim \frac{\sigma_A^2\sigma_B^2}{N_\text{rv}^2},
\end{equation}
where $\mathcal{N}_\text{eff} = \min \left( \mathcal{N}_A, \mathcal{N}_B \right)$, from which it follows that Eq.~\eqref{eq:var_P} goes as
\begin{equation}
    \text{Var}(P) \sim \frac{\sigma_A^2\sigma_B^2}{N_\text{rv}^2}\left[ 1 + \mathcal{O}\left(\frac{N_\text{rv}}{\mathcal{N}_\text{eff}}\right)\right].
\end{equation}
Therefore, the \gls*{RMS} error of $P$ will scale as $N_\text{rv}^{-1}$ when $\mathcal{N}_\text{eff} \gg N_\text{rv}$, but will cross over to $N_\text{rv}^{-1/2}$ scaling when $\mathcal{N}_\text{eff} \sim \mathcal{N}_\text{rv}$.
The relevant regime depends on the properties of the matrix elements of $G$, not just the vector dimension $V$. In particular, $N_\text{rv}\ll V$ does not guarantee $N_\text{rv}^{-1}$ scaling: if a row of $G$ is concentrated in the selected element, the single-vector fluctuations can be small and the linear terms can dominate. Nevertheless, for moderate $N_\text{rv}$, the additional computational cost associated with averaging over all $\binom{N_\text{rv}}{2}$ pairs of random vectors is worthwhile, as the dominant computational cost is still computing the $N_\text{rv}$ matrix-vector products $[G\xi_1,\dots,G\xi_{N_\text{rv}}]$. We typically suggest setting $N_\text{rv}\gtrsim 10$; for more information, refer to Ref.~\cite{Cohen-Stead2022Fast}.
}

\section{Applications and Performance}

We now provide benchmark simulation results for \revise{two prototypical models, the half-filled Holstein and \gls*{oSSH} models, each defined on a square lattice. Throughout, we perform \gls*{HQMC} simulations using the \gls*{smoqyelphqmc} package, and \gls*{DQMC} simulations using the \gls*{smoqydqmc} package for comparison.}

The Holstein model~\cite{Holstein1959Studies} is a simplified model for electrons interacting with the lattice vibrations of an Einstein solid. Its Hamiltonian is 
\begin{equation}\label{eq:holstein}
    \begin{aligned}
        \hat{H}_{\text{Hol.}} = & -t \sum_{\sigma, \mathbf{i}, \boldsymbol{\delta}}\left( \hat{c}^{\dagger}_{\sigma,\mathbf{i}+\boldsymbol{\delta}} \hat{c}^{\phantom\dagger}_{\sigma,\mathbf{i}} + \hat{c}^{\dagger}_{\sigma,\mathbf{i}} \hat{c}^{\phantom\dagger}_{\sigma,\mathbf{i}+\boldsymbol{\delta}}\right) - \mu \sum_{\sigma,\mathbf{i}} \hat{n}_{\sigma,\mathbf{i}} \\
        &+ \sum_{\mathbf{i}} \left( \frac{1}{2} M \Omega^2 \hat{X}_{\mathbf{i}}^2 + \frac{1}{2M}\hat{P}_{\mathbf{i}}^2\right)
        + \alpha_\text{h} \sum_{\sigma,\mathbf{i}}\hat{X}_{\mathbf{i}} \left( \hat{n}_{\sigma, \mathbf{i}} - \frac{1}{2}\right).
    \end{aligned}
\end{equation}
Here $\hat{c}^{\dagger}_{\sigma,\mathbf{i}} \ (\hat{c}^{\phantom \dagger}_{\sigma,\mathbf{i}})$ creates (annihilates) a spin-$\sigma$ electron on site $\mathbf{i}$ in the lattice; the sum over $\mathbf{i}$ runs over all lattice sites; $\boldsymbol{\delta} \in \{ \mathbf{x}, \mathbf{y} \}$ sums over the primitive square lattice vectors, \revise{with $a = |\boldsymbol{\delta}|$ the corresponding lattice constant}, such that the sum over \revise{$\mathbf{i}$} and $\boldsymbol{\delta}$ is over all nearest-neighbor sites; $t$ is the nearest-neighbor hopping integral; $\mu$ is the chemical potential with $\hat{n}_{\sigma, \mathbf{i}}^{\phantom\dagger} = \hat{c}^{\dagger}_{\sigma,\mathbf{i}} \hat{c}^{\phantom \dagger}_{\sigma,\mathbf{i}}$ the spin-$\sigma$ electron number operator for site $\mathbf{i}$. The third term in Eq.~\eqref{eq:holstein} describes the Einstein solid, with a single dispersionless phonon mode placed on each site in the lattice. The phonon position (momentum) operators are $\hat{X}_{\mathbf{i}} \ (\hat{P}_{\mathbf{i}})$, with an associated phonon frequency $\Omega$ and mass $M$. The fourth term in Eq.~\eqref{eq:holstein} couples the local electronic density to the phonon position, with the parameter $\alpha_\text{h}$ controlling the strength of the coupling. It is also common to define a dimensionless \gls*{eph} coupling constant $\lambda_\text{h} = \alpha_\text{h}^2/(M\Omega^2 W)$, where $W = 8t$ is the non-interacting bandwidth.

The \gls*{oSSH} model~\cite{Capone1997small} is another simplified model for \gls*{eph} coupling, whereby the relative atomic displacements modulate the electron hopping integrals. 
Its Hamiltonian is 
\begin{equation}\label{eq:oSSH}
    \begin{aligned}
        \hat{H}_\text{oSSH} = & -t \sum_{\sigma, \mathbf{i}, \boldsymbol{\delta}}\left( \hat{c}^{\dagger}_{\sigma,\mathbf{i}+\boldsymbol{\delta}} \hat{c}^{\phantom\dagger}_{\sigma,\mathbf{i}} + \hat{c}^{\dagger}_{\sigma,\mathbf{i}} \hat{c}^{\phantom\dagger}_{\sigma,\mathbf{i}+\boldsymbol{\delta}}\right) - \mu \sum_{\sigma,\mathbf{i}} \hat{n}_{\sigma,\mathbf{i}} \\
        & + \sum_{\mathbf{i},\boldsymbol{\delta}} \left( \frac{1}{2} M \Omega^2 \hat{X}_{\delta,\mathbf{i}}^2 + \frac{1}{2M}\hat{P}_{\delta,\mathbf{i}}^2\right) + \alpha_\text{o} \sum_{\sigma, \mathbf{i}, \boldsymbol{\delta}} \left[ \hat{X}_{\delta, \mathbf{i}+\boldsymbol{\delta}} - \hat{X}_{\delta, \mathbf{i}} \right] \left( \hat{c}^{\dagger}_{\sigma,\mathbf{i}+\boldsymbol{\delta}} \hat{c}^{\phantom\dagger}_{\sigma,\mathbf{i}} + \hat{c}^{\dagger}_{\sigma,\mathbf{i}} \hat{c}^{\phantom\dagger}_{\sigma,\mathbf{i}+\boldsymbol{\delta}}\right).
    \end{aligned}
\end{equation}
Most of the notation in Eq.~\eqref{eq:oSSH} is similar to that used for Eq.~\eqref{eq:holstein}. Here, however, two phonon modes are placed on each site $\mathbf{i}$ in the lattice that describe the motion of the atoms with $\hat{X}_{\delta,\mathbf{i}} \ (\hat{P}_{\delta,\mathbf{i}})$ denoting the position (momentum) operator of the atom on site $\mathbf{i}$ from its equilibrium position in the $\boldsymbol{\delta}$ direction. The fourth term in Eq.~\eqref{eq:oSSH} then introduces a \gls*{SSH}-like \gls*{eph} coupling mechanism whereby the relative displacement between adjacent atoms modulates the hopping amplitude between nearest-neighbor pairs of sites, with the strength of the interaction being controlled by the parameter $\alpha_\text{o}$. Once more, it is useful to define a dimensionless \gls*{eph} coupling strength $\lambda_\text{o} = (8\alpha_\text{o}^2)/(M\Omega^2 W)$~\cite{TanjaroonLy2023Comparative,TanjaroonLy2025Antiferromagnetic}.
\revise{For both the Holstein and \gls*{oSSH} models, we adopt units such that $\hbar = t = M = a = 1$, and set $\mu = 0$, which corresponds to half-filling $\frac{1}{N}\sum_{\mathbf{i},\sigma}\langle \hat{n}_{\mathbf{i},\sigma} \rangle = 1$.} Finally, the imaginary-time discretization is  set to $\Delta\tau = 0.05/t$ in all \gls*{QMC} simulations of both models.

\begin{figure}[t]
    \centering
    \includegraphics[width=\textwidth]{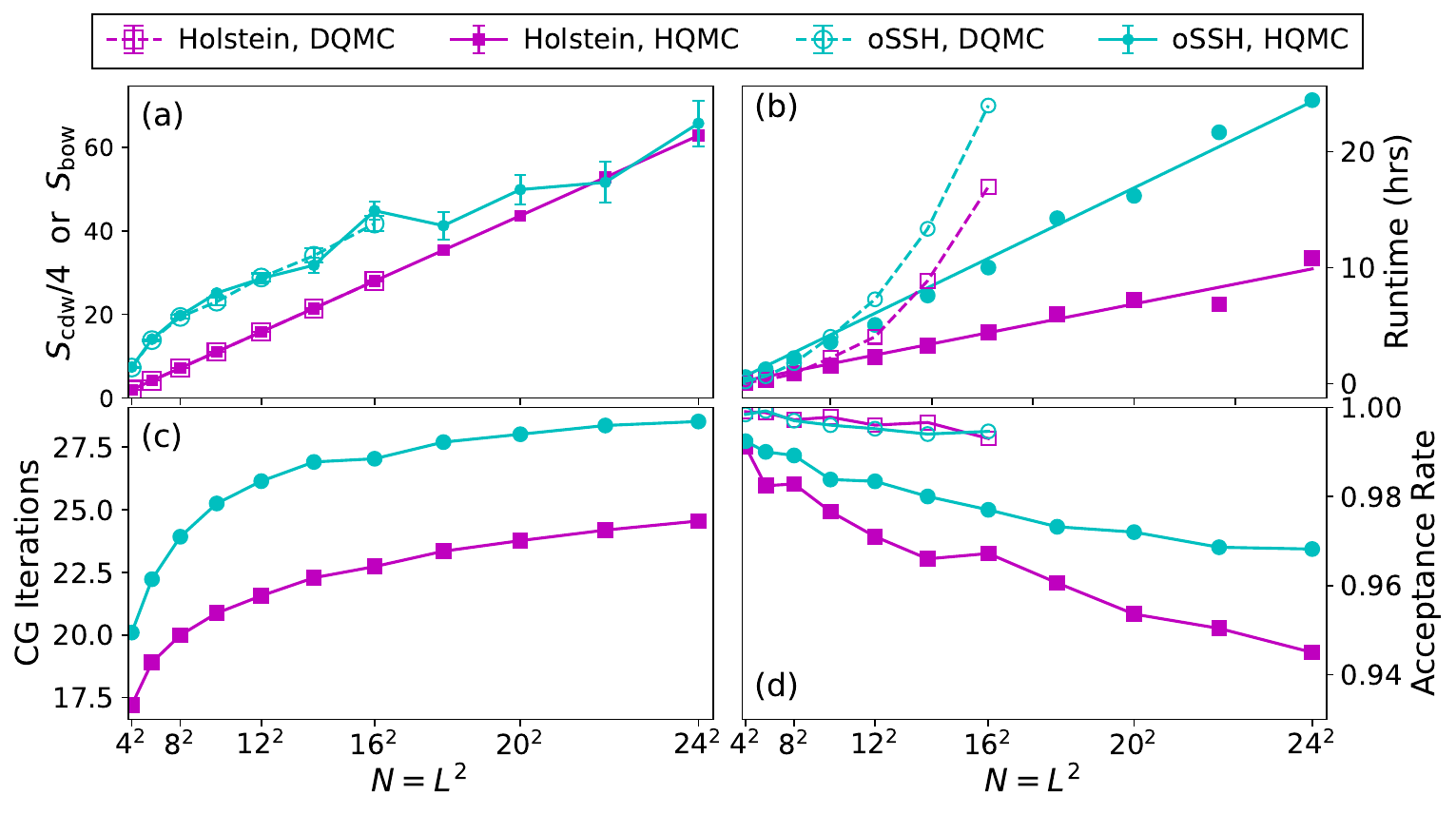}
    \caption{\revise{\gls*{HQMC} and \gls*{DQMC}} simulation results for the Holstein and \gls*{oSSH} models on square lattices of size $N = L^2$, for $\beta t = 10$, $\Omega/t = \revise{0.25}$, $\mu = 0$ (half-filling), and $\lambda_\text{h} = 0.25$ and $\lambda_\text{o} = \revise{0.16}$. Panel (a) displays the \gls*{CDW} and \gls*{BOW} structure factors measured in simulations of the Holstein and \gls*{oSSH} model simulations, respectively. Panel (b) displays the runtime for simulations of each model (see text for details of the hardware). \revise{The solid lines are a linear guide for the eye for the \gls*{HQMC} runtime results.} Panel (c) displays the average number of \gls*{CG} iterations per solve during the \revise{\gls*{HQMC}} simulation\revise{s}. Panel (d) displays the acceptance rate for the \gls*{HMC} updates in the simulations.}
    \label{fig:runtime_vs_L}
\end{figure}
\begin{figure}[t]
    \centering
    \includegraphics[width=\textwidth]{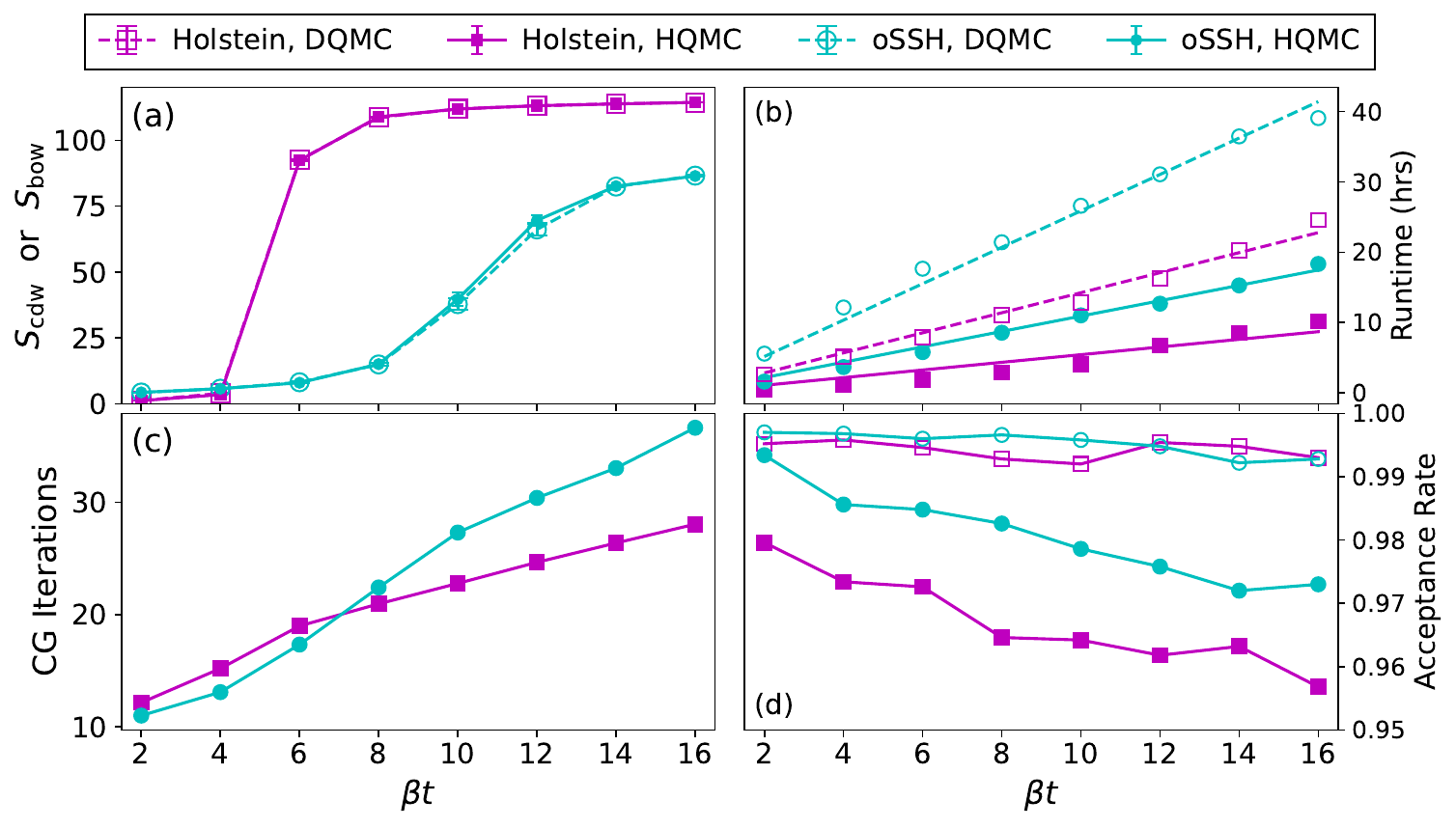}
    \caption{\revise{\gls*{HQMC} and \gls*{DQMC} simulation results for the} Holstein and \gls*{oSSH} models on a $N = 16 \times 16$ site square lattice at various inverse temperatures $\beta$, with $\Omega/t = \revise{0.25}$ and $\mu = 0.0$, and $\lambda_\text{h} = 0.25$ and $\lambda_\text{o} = \revise{0.16}$ in each model, respectively. Panel (a) displays the \gls*{CDW} and \gls*{BOW} structure factors measured in simulations of the Holstein and \gls*{oSSH} model simulations, respectively. Panel (b) displays the runtime for simulations of each model. \revise{The dashed and solid lines are linear fits serving as a guide for the eye.} Panel (c) displays the average number of \gls*{CG} iterations per solve during the \revise{\gls*{HQMC} simulations}, averaged over both updates and measurements. Panel (d) displays the acceptance rate for the \gls*{HMC} updates in the simulations.}
    \label{fig:runtime_vs_beta}
\end{figure}

The half-filled Holstein model on a square lattice undergoes a finite-temperature transition into an ordered $\mathbf{Q} = (\pi,\pi)$ \gls*{CDW} phase~\cite{Scalettar1989Competition, Dee2019temperature, Costa2020phase, Cohen-Stead2019Effect,Costa2017Principal, Costa2018Phonon, Weber2018Twodimensional}, in which the electrons preferentially localize on a single sub-lattice. This phase transition is in the \gls*{2D} Ising universality class, and can be detected by measuring $S_\text{cdw} = S_C(\mathbf{Q})$, where
\begin{equation}
    S_C(\mathbf{q}) = \frac{1}{N}\sum_{\mathbf{i},\mathbf{r}} e^{-i\mathbf{q}\cdot\mathbf{r}} \langle \hat{n}_{\mathbf{i}+\mathbf{r}} \hat{n}_\mathbf{i} \rangle
\end{equation}
is the momentum-resolved charge structure factor.
Similarly, at weak coupling, the half-filled \gls*{oSSH} model has a finite-temperature phase transition to an insulating $\mathbf{Q} = (\pi,\pi)$ \gls*{BOW} phase that breaks a local $C_4$ rotation symmetry~\cite{TanjaroonLy2025Antiferromagnetic, silva2025twodimensional}. \revise{Consistent with previous work\revise{~\cite{Cai2021antiferromagnetism,Gotz2022valence}}, the} strength of the \gls*{BOW} correlations \revise{is quantified using} the bond-structure factor 
\begin{equation}
S_\text{bow} = S_{B,(x,x)}(\mathbf{Q}) + S_{B,(y,y)}(\mathbf{Q}), 
\end{equation}
where
\begin{equation}
    S_{B,(\delta',\delta)}(\mathbf{q}) = \frac{1}{N} \sum_{\mathbf{i},\mathbf{r}} e^{-i\mathbf{q}\cdot\mathbf{r}} \langle \hat{B}_{\mathbf{i}+\mathbf{r},\boldsymbol{\delta}'}  \hat{B}_{\mathbf{i},\boldsymbol{\delta}} \rangle
\end{equation}
is the momentum-resolved bond structure factor, and
\begin{equation}
    \hat{B}_{\mathbf{i}, \boldsymbol{\delta}} = \revise{\sum_{\sigma}} \left( \hat{c}_{\sigma,\mathbf{i}+\boldsymbol{\delta}}^{\dagger} \hat{c}_{\sigma,\mathbf{i}}^{\phantom \dagger} + \hat{c}_{\sigma,\mathbf{i}}^{\dagger} \hat{c}_{\sigma,\mathbf{i}+\boldsymbol{\delta}}^{\phantom\dagger}\right)
\end{equation}
is the bond operator. \revise{Note that the $S_{B,(x,y)}(\mathbf{q})$ and $S_{B,(y,x)}(\mathbf{q})$ cross terms have been neglected as they vanish in the thermodynamic limit \revise{for the ``standard'' \gls*{BOW} phase studied below. They should be included in the general case, e.g., for staircase orders like those reported in Ref.~\cite{silva2025twodimensional}.} 

Lastly, we also measure the s-wave pair susceptibility
\begin{equation}
    P_s = \frac{1}{N} \int_0^\beta \langle \hat{\Delta}(\tau) \hat{\Delta}^\dagger(0)\rangle \ d\tau,
\end{equation}
where $\hat{\Delta}(\tau) = \sum_\mathbf{i} \hat{c}_{\downarrow,\mathbf{i}}(\tau) \hat{c}_{\uparrow,\mathbf{i}}(\tau)$.}
\begin{figure}[t]
    \centering
    \includegraphics[width=1.0\linewidth]{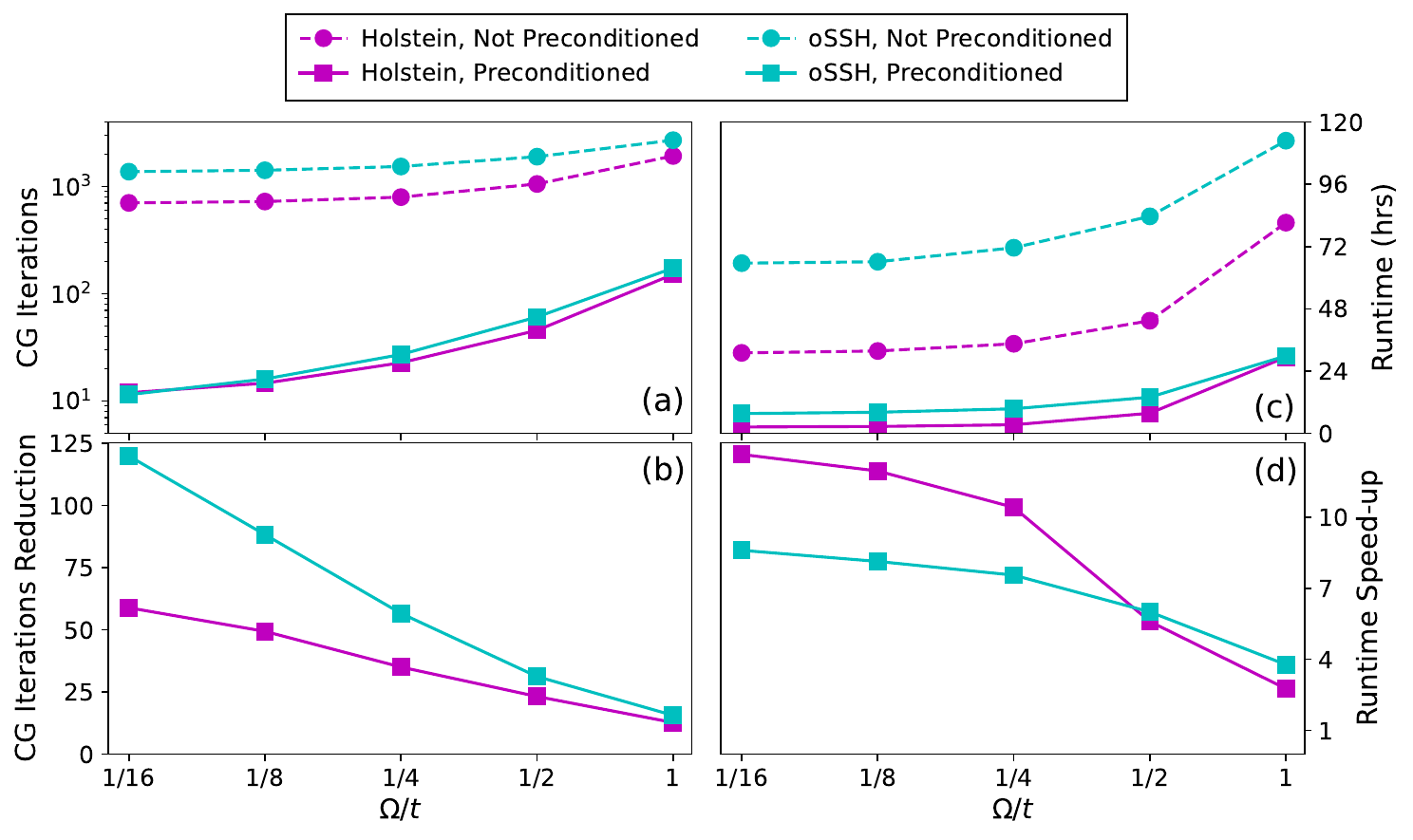}
    \caption{\revise{Results for the Holstein and \gls*{oSSH} models versus phonon frequency $\Omega/t$ on a $N = 16 \times 16$ lattice at $\beta t = 10$ and $\mu=0$, with $\lambda_\text{h} = 0.25$ and $\lambda_\text{o} = 0.16$ for each model, respectively. Panel (a) displays the number of \gls*{CG} iterations for convergence in the simulations of each model both with and without the preconditioner being used. Panel (b) then plots the relative factor by the CG iterations are reduced when the preconditioner is introduced. Panel (c) is then the corresponding runtime of each simulation in hours, with panel (d) then displaying the relative speed-up that results from using the preconditioner.}}
    \label{fig:freq_sweep}
\end{figure}
\begin{figure}[t]
    \centering
    \includegraphics[width=\textwidth]{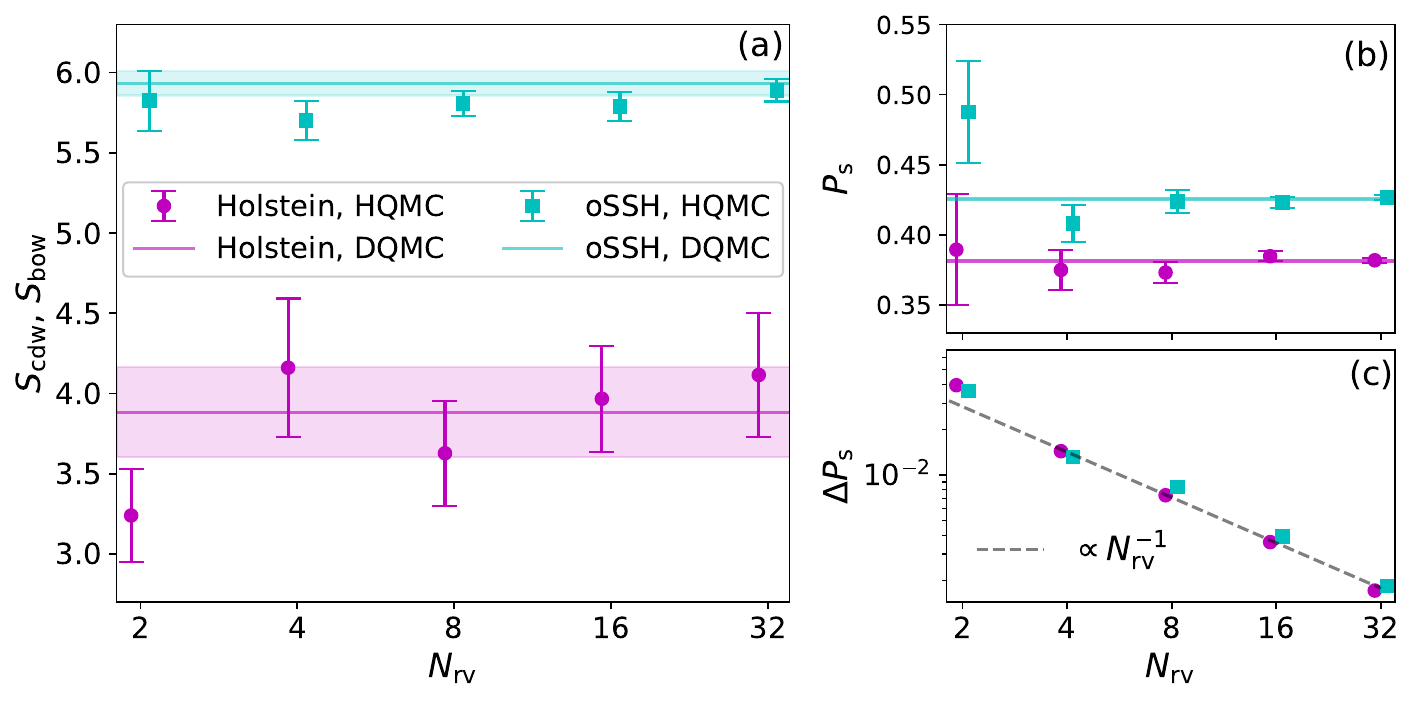}
    \caption{\revise{Reported values for relevant observables in simulations of the Holstein and \gls*{oSSH} model using different numbers of random vectors $N_\text{rv}$ to make measurements. Comparison is made against \gls*{DQMC} calculations performed using the \gls*{smoqydqmc} package. Results are for an $L = 16$ lattice at $\beta t = 4$ with a phonon frequency $\Omega/t = 0.5$, and $\lambda_\text{h} = 0.25$ and $\lambda_\text{o} = 0.16$ for each model respectively. Panel (a) displays the \gls*{CDW} and \gls*{BOW} structure factors as a function of $N_\text{rv}$ in simulations of the Holstein and \gls*{oSSH} models, respectively. Panel (b) reports the measured pair susceptibility for each model, and panel (c) shows how the corresponding error bars approximately fall off as $N_\text{rv}^{-1}$. All error bars are $1\sigma$ standard errors obtained using a jackknife procedure.}}
    \label{fig:Nrv_fig}
\end{figure}
\revise{
\begin{figure}[t]
    \centering
    \includegraphics[width=1.0\linewidth]{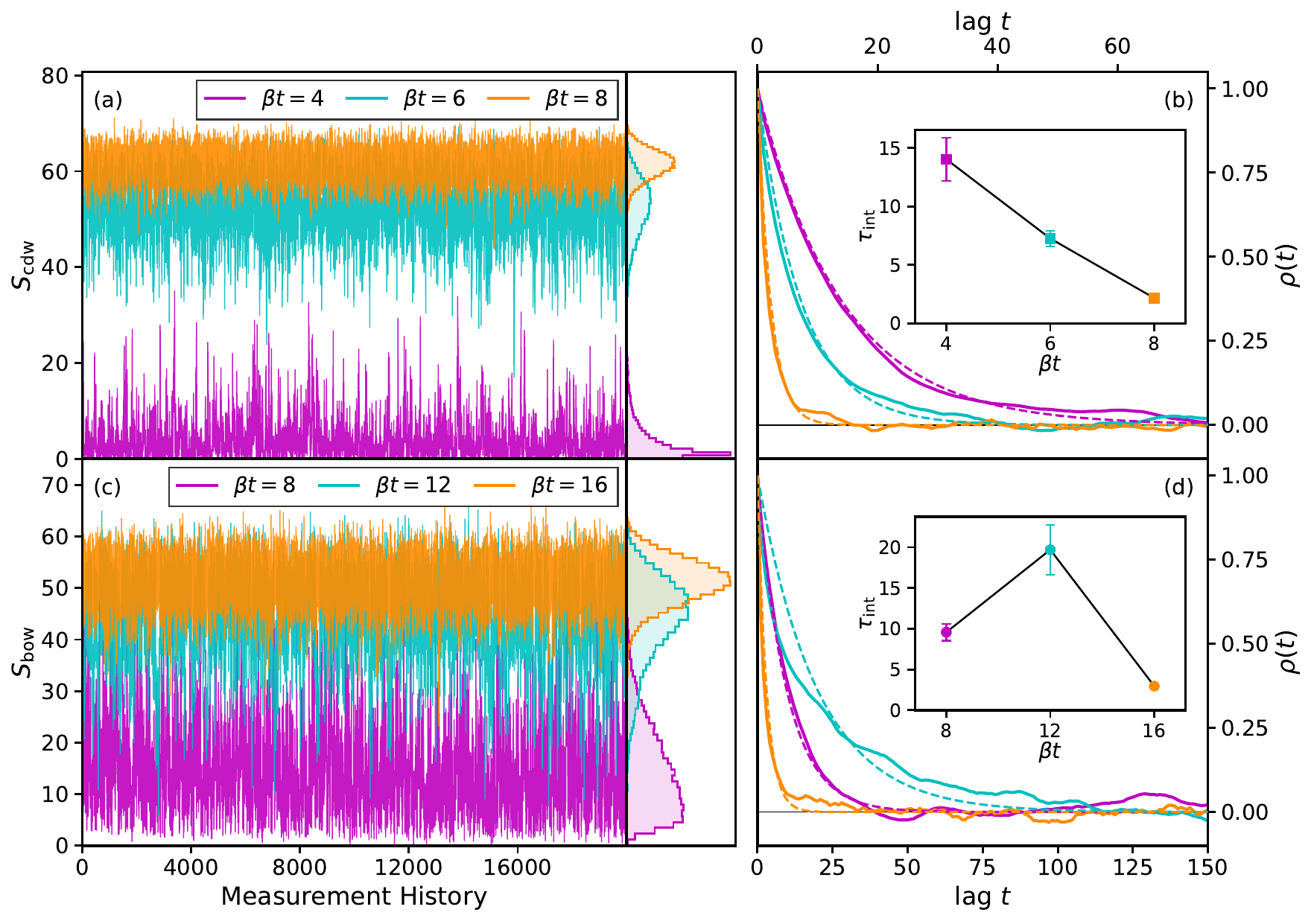}
    \caption{Autocorrelation time results in simulations of the Holstein and \gls*{oSSH} models on a $L = 12$ lattice for $\Omega/t = 0.25$, with dimensionless couplings $\lambda_\text{h} = 0.25$ and $\lambda_\text{o} = 0.16$, respectively. Panel (a) and (c) show is the histories of sampled $S_\text{cdw}$ and $S_\text{bow}$ values from simulations of the Holstein and \gls*{oSSH} models at three representative temperatures, respectively, with histograms of those sampled values shown vertically on the right side of the panel. Panels (b) and (d) are plots of the corresponding normalized autocorrelation $\rho(t)$ versus lag $t$, with the dashed lines showing fits based on the measured integrated autocorrelation times $\tau_\text{int}$ shown in the insets.}
    \label{fig:autocorrelation}
\end{figure}
}
\begin{figure}[t]
    \centering
    \includegraphics[width=\textwidth]{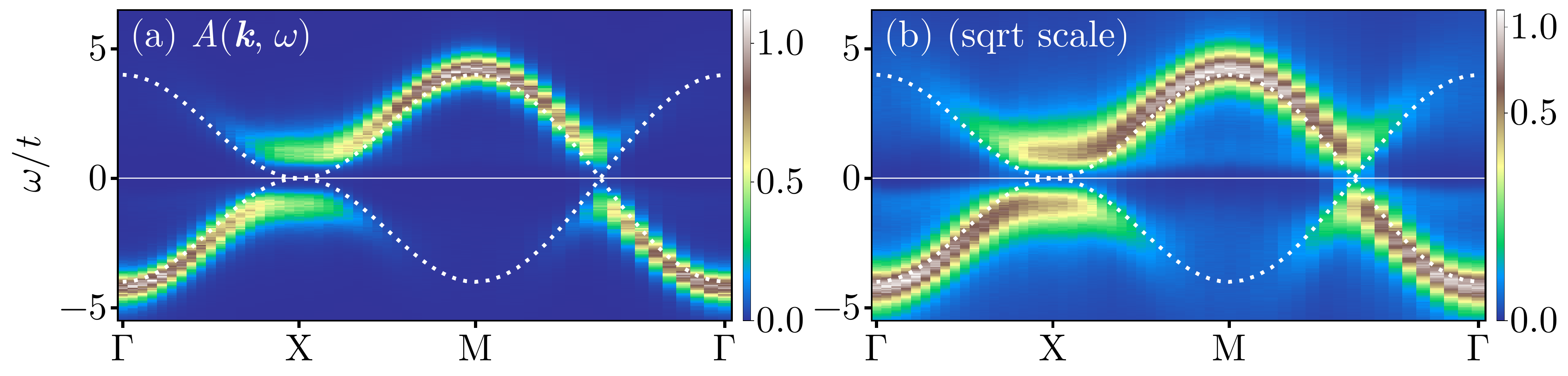}
    \caption{The Holstein model single-particle electron spectral function $A(\mathbf{k},\omega)$ as a function of momentum $\mathbf{k}$ along a \revise{high-symmetry} cut through the \gls*{BZ}. Solid white lines at $\omega = E_\text{F} = 0$ indicate the Fermi energy. The dashed white lines denote the non-interacting electron band, and its reflection about the Fermi energy $E_\text{F} $. All results are for a $N = 36 \times 36$ square lattice at $\beta t=10.0$ and phonon energies of $\Omega/t=0.5$. (a) The Holstein model spectral function $A(\mathbf{k},\omega)$ for $\lambda_\text{h}=0.25$. (b) Holstein model spectral function plotted on a square-root scale to accentuate finer features.}
    \label{fig:spectral_function_holstein}
\end{figure}
\begin{figure}[t]
    \centering
    \includegraphics[width=\textwidth]{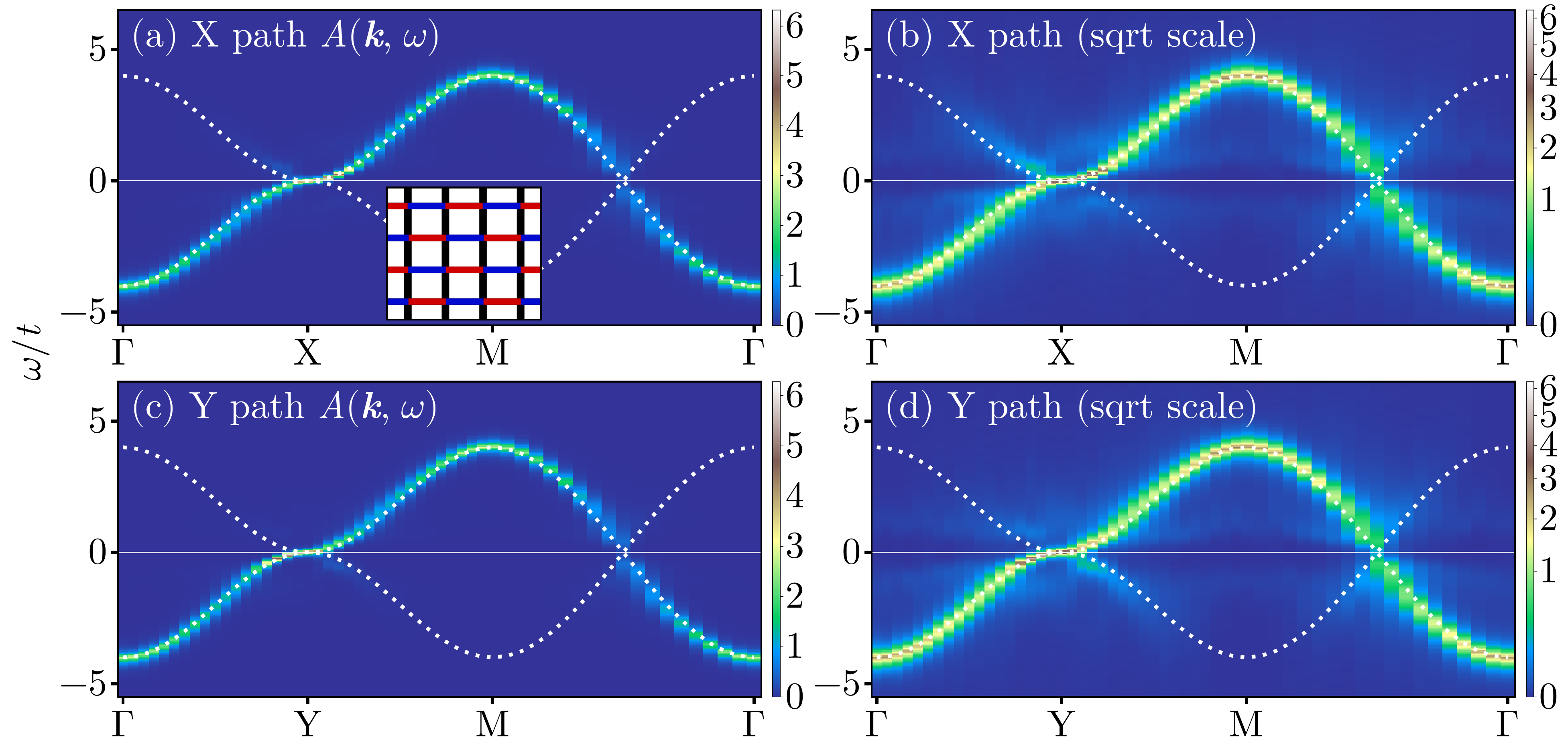}
    \caption{The oSSH model single-particle electron spectral function $A(\mathbf{k},\omega)$ as a function of momentum $\mathbf{k}$ along a \revise{high-symmetry cut} through the \gls*{BZ}. Solid white lines at \revise{$\omega = E_\mathrm{F} = 0$} indicate the Fermi energy. The dashed white lines denote the non-interacting electron band, and its reflection about the Fermi energy \revise{$E_\mathrm{F}$}. All results are for a $N = 36 \times 36$ square lattice at $\beta t=10.0$ and phonon energies of $\Omega/t=0.5$. (a) The oSSH model spectral function $A(\mathbf{k},\omega)$ for $\lambda_\text{o}=\revise{0.16}$ along the $\Gamma-X-M-\Gamma$ path. The inset shows an $\hat{x}$-oriented BOW phase. (b) X-oriented path spectral function plotted on a square-root scale to accentuate finer features. (c) The oSSH model spectral function $A(\mathbf{k},\omega)$ for $\lambda_\text{o}=\revise{0.16}$ along the $\Gamma-Y-M-\Gamma$ path.  (d) Y-oriented path spectral function plotted on a square-root scale. }
    \label{fig:spectral_function_oSSH}
\end{figure}

Figures~\ref{fig:runtime_vs_L}~and~\ref{fig:runtime_vs_beta}  report simulation results for both models; Fig.~\ref{fig:runtime_vs_L} shows results as a function of system size $N = L \times L$ with fixed $\beta$ while Fig.~\ref{fig:runtime_vs_beta} shows results \revise{as} a function of $\beta$ and fixed system size. When performing these benchmarking tests, all Holstein and \gls*{oSSH} model simulations ran on Cascade Lake nodes of Intel Xeon Gold 6248R processors. \revise{Both the \gls*{HQMC} and \gls*{DQMC} simulations} performed 1,000 thermalization \gls*{EFA-HQMC} \revise{and radial} updates, followed by an additional 4,000 iterations of alternating \revise{the} \gls*{EFA-HQMC} \revise{and radial} updates \revise{with} measurements. Each \gls*{EFA-HQMC} update consisted of $N_t = 16$ time-steps with an average integrated trajectory time of $T_t = \pi/2$ such that the time-step is given by \revise{$\Delta t = T_t/N_t$}. Note that the time-step $\Delta t$ and integrated trajectory \revise{time $T_t$ were} also randomly jittered by 5\% at the start of each \gls*{EFA-HQMC} update~\cite{Cohen-Stead2022Fast, Mackenze1989Improved}. A tolerance of $\delta = 1 \times 10^{-10}$ was used when performing \gls*{CG} solves for accepting or rejecting an update or making measurements, while a tolerance of $\delta = 1 \times 10^{-5}$ was used when evaluating the forces in a \gls*{EFA-HQMC} update. Note that the average \gls*{EFA-HQMC} acceptance rate, as reported in Figs.~\ref{fig:runtime_vs_L}(d) and \ref{fig:runtime_vs_beta}(d), remained above 90\% in all simulations. \revise{Additionally}, $N_\text{rv} = 10$ random vectors were used and averaged over each time measurements were made. \revise{Lastly, all Holstein and \gls*{oSSH} model simulations were initialized from phonon configurations corresponding to \gls*{CDW} and \gls*{BOW} order, respectively, to reduce thermalization times.}

Figures~\ref{fig:runtime_vs_L}(a)~and~\ref{fig:runtime_vs_beta}(a) show that $S_\text{cdw}$ and $S_\text{bow}$ increase monotonically with both $N$ and $\beta$, respectively, consistent with the formation of \gls*{CDW} and \gls*{BOW} order in each model at low temperatures. Additionally, the significant jump observed in $S_\text{cdw}$ in Fig.~\ref{fig:runtime_vs_beta}(a) at $\beta t \sim 6$ is consistent with previous work that reported a transition temperature of $T_c/t \approx 1/6$ for a $\lambda_\text{h} = 0.25$ Holstein model~\cite{Cohen-Stead2019Effect,Costa2017Principal, Costa2018Phonon, Weber2018Twodimensional}. This explains why $S_\text{cdw}$ increases linearly with $N$ in Fig.~\ref{fig:runtime_vs_L}(a), as the temperature $T/t = 1/10$ is significantly below the transition temperature. However, the $S_\text{bow}$ curve in Fig.~\ref{fig:runtime_vs_beta}(a) suggests that this temperature is close to but above the \gls*{BOW} transition temperature, which would also explain why $S_\text{bow}$ increases in a sub-linear fashion with $N$ in Fig.~\ref{fig:runtime_vs_L}(a). A more precise finite size scaling analysis to determine the \gls*{BOW} transition temperature is beyond the scope of this work.
Next, the runtime for simulations of both models as a function of system size $N$ and inverse temperature $\beta$ is reported in Figs.~\ref{fig:runtime_vs_L}(b)~and~\ref{fig:runtime_vs_beta}(b), respectively. Both sets of \revise{\gls*{HQMC}} simulations \revise{clearly demonstrate better-than-cubic scaling with system size}. \revise{This behavior} can be traced to the number of \gls*{CG} iterations per solve, which \revise{clearly scales sub-quadratically} with $N$ and $\beta$, as shown in Figs.~\ref{fig:runtime_vs_L}(c) and \ref{fig:runtime_vs_beta}(c), respectively. As these simulations use the preconditioning method introduced in Sec.~\ref{sec:kpm_preconditioner}, which is based on an adiabatic approximation, we expect the number of \gls*{CG} iterations to \revise{become approximately} independent of $N$ with decreasing $\Omega$, as reported in Ref.~\cite{Cohen-Stead2022Fast}.

\revise{These expectations are consistent with Fig.~\ref{fig:freq_sweep}, which reports benchmarking simulation results over a range of phonon frequencies $\Omega$ for both the Holstein and \gls*{oSSH} models, both with and without the preconditioner being applied in the simulations. We see that in both cases, the number of \gls*{CG} iterations and runtime increase monotonically with the phonon frequency. Additionally, while the preconditioner accelerates the simulations in all cases, the relative speed-up and reduction in \gls{CG} iterations becomes larger as the phonon frequency is decreased, and the adiabatic limit is approached.}

\revise{Next, to demonstrate the efficacy of the stochastic measurement methods introduced in Sec.~\ref{sec:stoch_meas}, Fig.~\ref{fig:Nrv_fig} displays correlations measurements from simulations of the Holstein and \gls*{oSSH} models versus the number of random vectors $N_\text{rv}$ used to make measurements. The results are also compared against comparable \gls*{DQMC} simulations performed using \gls*{smoqydqmc}~\cite{Cohen-Stead2024SmoQyDQMCjl}. There are two sources of noise in the reported measurement uncertainty, one originating from the Monte Carlo sampling process and another originating from the stochastic nature of the measurement itself. In Fig.~\ref{fig:Nrv_fig}(a) we see that the uncertainty in the measured \gls*{CDW} structure factor in the Holstein model simulations appears approximately independent of $N_\text{rv}$, and is consistent with the uncertainty reported by the \gls*{DQMC} simulation. This suggests that the reported error is a result of sampling related fluctuations. In the case of the \gls*{BOW} structure factor measured in the \gls*{oSSH} model simulations, the reported uncertainty subtly decreases out to $N_\text{rv} = 8$, beyond which it appears to saturate to a value consistent with the \gls*{DQMC} result. This behavior suggests that the residual uncertainty for $N_\text{rv} \ge 8$ once more results from sampling.}

\revise{It makes sense that both the uncertainty in \gls*{CDW} and \gls*{BOW} structure factors would be limited by the sampling, as both are sensitive to large sampling fluctuations associated with the emergent order that arises in each model. If we instead turn our attention to the pair susceptibility measurement shown in Fig.~\ref{fig:Nrv_fig}(b), we see that the reported uncertainty strongly depends on $N_\text{rv}$, indicating that sampling fluctuations are small as the pairing correlations in each system are suppressed by the emergent \gls*{CDW} and \gls*{BOW} phases. As such, the uncertainty $\Delta P_s$ in the pair susceptibility measurement is primarily a function of the stochastic nature of the measurement process. This interpretation is corroborated by Fig.~\ref{fig:Nrv_fig}(c), which clearly shows the uncertainty in the pair susceptibility falling off with the expected $\Delta P_s \propto N_\text{rv}^{-1}$ behavior described in Sec.~\ref{sec:reducing_stoch_errors}}.

\revise{As an additional comment, as in any Monte Carlo simulation, the total statistical uncertainty depends on the autocorrelation time of the sampling process. The autocorrelation time of \gls*{smoqyelphqmc} simulations can vary significantly between different \gls*{eph} models, and even for a single model as a function of coupling strength, doping, and temperature. To provide some limited insight into how the integrated autocorrelation time may vary between models, and as a function of temperature, Fig.~\ref{fig:autocorrelation} shows sampling histories and corresponding autocorrelation times for the $S_\text{cdw}$ and $S_\text{bow}$ structure factors in representative simulations of the Holstein and \gls*{oSSH} models. We see that the reported autocorrelation times resulting from measurements separated by an \gls*{EFA-HQMC} and radial update can vary significantly, highlighting that sampling efficiency strongly depends on the details of the model. Therefore, it remains important to assess sampling efficiency when interpreting results generated using \gls*{smoqyelphqmc}, and we caution the reader against over-generalizing the integrated autocorrelation times reported in Fig.~\ref{fig:autocorrelation} to other scenarios. Further discussion of sampling efficiency and autocorrelation times when \gls*{HMC} and similar methods are used to simulate \gls*{eph} models can be found in Refs.~\cite{Cohen-Stead2022Fast, Batrouni2019Langevin, Ostmeyer2025Minimal, Ostmeyer2025Exponential}.}

\revise{Lastly}, to \revise{further} demonstrate the utility of the \gls*{smoqyelphqmc} package, we ran additional simulations of both models on a larger $N = 36 \times 36$ lattice at $\beta t = 10$. These simulations ran 30 walkers in parallel, thermalized the system using 2,000 \gls*{EFA-HQMC} updates, and then performed an additional 2,500 updates, after each of which measurements were made. \revise{The simulations} were run on a single Ryzen AI Max+ 395 processor with 128GB LPDDR5X-8000MT/s memory, with each walker finishing in approximately 4.5 days. Measurements were further rebinned down to 20 bins per walker. We then analytically continued the measured momentum-space single-particle electron Green's function to the real axis by inverting the integral equation 
\begin{equation}
    G_\sigma(\mathbf{k},\tau) = \langle \hat{c}_{\mathbf{k},\sigma}^{\phantom\dagger}(\tau) \hat{c}_{\mathbf{k},\sigma}^{\dagger}(0) \rangle = \int_{-\infty}^{+\infty} d\omega \ \frac{e^{-\tau\omega} }{1 + e^{-\beta\omega}} \ A(\mathbf{k},\omega),
\end{equation}
where $A(\mathbf{k},\omega)$ is the single-particle spectral function. This task was performed using the \gls*{DEAC} algorithm~\cite{nichols_2022_parameter}, as implemented in the \gls*{smoqydeac} package~\cite{Neuhaus2024SmoQyDEACjl}. For each $\mathbf{k}$ point, \gls*{smoqydeac} conducted 2,000 independent runs averaged together with an output over 601 evenly spaced energies $\omega/t\in[-15,15]$. The complete \revise{Green's} function covariance matrix was used for each $\mathbf{k}$ point, helping to resolve more subtle spectral features~\cite{jarrell_1996_bayesian,Neuhaus2024SmoQyDEACjl}. The resulting spectral functions for the Holstein and \gls*{oSSH} models are shown in \revise{Figs.~\ref{fig:spectral_function_holstein} and \ref{fig:spectral_function_oSSH}}, respectively.

\revise{Figures~\ref{fig:spectral_function_holstein}(a) and \ref{fig:spectral_function_holstein}(b)  plot the spectral function $A(\mathbf{k},\omega)$ for the Holstein model} along a high-symmetry cut through the \gls*{BZ} on linear and square-root intensity scales, respectively. The spectral function exhibits clear signatures of insulating \gls*{CDW} order, including a gap at the Fermi surface and visible back-folding of spectral weight between the $\Gamma$ and $M$ points, as expected for \revise{$\mathbf{Q}=(\pi,\pi)$} ordering.

\revise{Corresponding results for the \gls*{oSSH} spectral function are shown in .~\ref{fig:spectral_function_oSSH}(a)-(d). In order to bias the directional preference of the emergent \gls*{BOW} phase, we introduced weakly anisotropic \gls*{eph} couplings, such that in the $\hat{x}$ and $\hat{y}$ directions the microscopic coupling strengths \revise{are} $\alpha_{\text{o},x} = 0.206$ and $\alpha_{\text{o},y} = 0.194$, respectively. Panel (a) shows the high-symmetry cut through the \gls*{BZ} containing the X \revise{high-symmetry} point. The inset is a position-space representation of the expected $\hat{x}$-oriented \gls*{BOW} phase, where the black bonds are undistorted, and the blue and red bonds are alternately expanded and contracted. Panel (b) is the same $X$-oriented high-symmetry cut plotted with a square-root scale. In Fig.~\ref{fig:spectral_function_oSSH}(c) we have instead plotted the high-symmetry cut through the \gls*{BZ} containing the Y \revise{high-symmetry} point, with the corresponding square-root scale plot shown in panel (d). For both we observe subtle backfolding features at the $X$ and $Y$ points, consist with the formation of \gls*{BOW} correlations in the $x$-oriented direction.}

\section{Summary and Conclusions}
The \gls*{smoqyelphqmc} package extends the functionality of the \gls*{smoqydqmc} package, enabling \gls*{QMC} simulations of real, spin-symmetric, \gls*{eph} Hamiltonians with a computational cost that scales near-linearly in both the system size $N$ and inverse temperature $\beta$. More specifically, this package exports an open-source implementation of the \gls*{QMC} method described in Ref.~\cite{Cohen-Stead2022Fast}, with additional algorithmic improvements and support for a broad class of \gls*{eph} models. The \gls*{smoqyelphqmc} \revise{package} therefore enables nonperturbative simulation of \gls*{eph} Hamiltonians defined on significantly larger lattices than is possible with the traditional \gls*{DQMC} method. This is all done while retaining the flexible scripting interface first introduced in the \gls*{smoqydqmc} package. Additionally, users will find that just as is the case \revise{for the \gls*{smoqydqmc} package}, there is extensive \href{https://smoqysuite.github.io/SmoQyElPhQMC.jl/dev/}{online documentation} for \revise{the \gls*{smoqyelphqmc} package} that includes a frequently updated list of tutorials and examples that users can use to  learn how to use the package.

\section*{Acknowledgments} 
We thank R. T. Scalettar for useful discussions and collaborations \revise{on} the algorithms utilized by this package.

\paragraph{Funding information:} The implementation of the \gls*{smoqyelphqmc} package and its documentation was supported by the National Science Foundation under grant No. OAC-2410280. Many of the algorithms that have been implemented in this package were initially developed with support from the U.S. Department of Energy, Office of Science under grant No. DE-SC0022311. B.C.-S. acknowledges additional support from the Simons Foundation for the continued maintenance of the SmoQy Suite. 


\bibliography{references_manual}

\end{document}